\documentclass[a4paper,10pt]{article}
\usepackage{amsfonts}
\usepackage[dvips]{graphicx}
\usepackage{graphpap}
\usepackage{ifthen}
\usepackage{amsmath}
\usepackage{amssymb}
\usepackage{amscd,latexsym}
\usepackage{euscript}
\include{psfig}
\usepackage{theorem}
\newfam\scrfam
\batchmode\font\tenscr=rsfs10 \errorstopmode
\ifx\tenscr\nullfont
\else   \font\sevenscr=rsfs7 
        \font\fivescr=rsfs5 
        \skewchar\tenscr='177 \skewchar\sevenscr='177 \skewchar\fivescr='177
        \textfont\scrfam=\tenscr \scriptfont\scrfam=\sevenscr
        \scriptscriptfont\scrfam=\fivescr
        \def\scr{\fam\scrfam}
\fi
\newcommand{\eq}[1]{\begin{equation}#1\end{equation}}              %for single equation with eqn #
\newcommand{\eqnono}[1]{\begin{displaymath}#1\end{displaymath}}    %for single equation without eqn #
\newcommand{\eqsnono}[1]{\begin{eqnarray*}#1\end{eqnarray*}}       %for multi line equations, or eqns with matrices etc, without eqn #
\newcommand{\mx}[2]{\left(\begin{array}{#1}#2\end{array}\right)}   %matrix command
\newcommand\rctr{\renewcommand{\theenumi}{(\roman{enumi})}}

\newcommand{\rnum}[1]{\rctr\begin{enumerate}#1\end{enumerate}}

\newcommand{\eqal}[2]{\begin{alignat*}{#1}#2\end{alignat*}}
\def\a{\alpha}					
\def\b{\beta}					
					
\def\d{\delta}

		\def\gog{{\mathfrak g}}

\def\k{\kappa}					
					
\def\l{\lambda}			\def\scL{{\scr L}}		\def\clL{{\cal L}}
\def\m{\mu}					
					
\def\r{\rho}		\def\goo{{\mathfrak o}}			
					\def\clP{{\cal P}}
\def\s{\sigma}					
				\def\dsR{{\mathbb R}}	
					
			\def\scT{{\scr T}}		
					
\def\o{\omega}

\def\O{\Omega}					
\def\L{\Lambda}

\def\vf{\varphi}
\newcommand\ve{\varepsilon}

\newcommand\esD{{\EuScript D}}

\newcommand\esF{{\EuScript F}}

\newcommand\Lie{\scL}
\newcommand\flie{\hat{\scL}}
\newcommand\vlie{\check{\scL}}
\newcommand\fr{\over}
\newcommand\pa{\partial}
\newcommand\ri{{}^\sharp\!}
\newcommand\li{{}^\flat\!}
\newcommand\ct{{}^c\!}
\newcommand{\no}[1]{\thinspace\overset{{}^{#1}\!\surd}{\ldots}\thinspace}

\newcommand{\del}{\nabla}

\newcommand{\minus}{\!{\raise1pt\hbox{$\scriptstyle{-}$}}\!}

\newcommand{\fraction}[1]{{\lower2.5pt\hbox{\eightrm 1}\/\raise2.5pt\hbox{\eightrm #1}}}
\newcommand{\Fraction}[2]{{\lower2.5pt\hbox{\eightrm #1}\/\raise2.5pt\hbox{\eightrm #2}}}
\newcommand{\tr}{\hbox{\rm tr}}

\newcommand{\ul}{\underline}

\theoremstyle{break}
\newtheorem{theorem}{Theorem}[section]
\newtheorem{lemma}[theorem]{Lemma}
\newtheorem{conjecture}[theorem]{Conjecture}
\newtheorem{proposition}[theorem]{Proposition}
\newtheorem{corollary}[theorem]{Corollary}
\newtheorem{definition}[theorem]{Definition}
\newtheorem{example}[theorem]{Example}
\newtheorem{remark}[theorem]{Remark}
\newtheorem{notation}[theorem]{Notation}
\newtheorem{convention}[theorem]{Convention}

\newcommand{\thm}[2]{\begin{theorem}#2\label{#1}\end{theorem}}
\newcommand{\lem}[2]{\begin{lemma}#2\label{#1}\end{lemma}}

\newcommand{\prop}[2]{\begin{proposition}#2\label{#1}\end{proposition}}
\newcommand{\cor}[2]{\begin{corollary}#2\label{#1}\end{corollary}}
\newcommand{\defn}[2]{\begin{definition}#2\label{#1}\end{definition}}
\newcommand{\remk}[2]{\begin{remark}#2\label{#1}\end{remark}}
\newcommand{\nota}[2]{\begin{notation}#2\label{#1}\end{notation}}
		
\newcommand{\exam}[2]{\vskip .1cm\noindent\rule{12.2cm}{.5pt}{\begin{example}\vskip -.4cm\begin{list}{}{\leftmargin 1cm\labelsep .7cm}\item {}\item{\small #2}
	\end{list}\end{example}}\label{#1}\vskip -.5cm\noindent\rule{12.2cm}{.5pt}\vspace{.4cm}}
\newcommand{\proof}[1]{{\em \small proof:}\vspace{-19pt}\begin{list}{}{\leftmargin 1.2cm\labelsep .7cm}\item{\small #1}\hfill$\blacksquare$\end{list}}

\newcommand{\rchapter}[1]{\newpage\label{dummy#1}
	\ifthenelse{\isodd{\pageref{dummy#1}}}{}{\blankpage}
	\chapter{#1}}

\newcommand{\rchaptno}[1]{\newpage\label{dummy#1}
	\ifthenelse{\isodd{\pageref{dummy#1}}}{}{\blankpage}
	\chapter*{#1}}

\newcommand{\ws}{{\scr M}}	\newcommand{\ts}{\ul{\scr M}}
		\newcommand{\tg}{\ul{g}}	\newcommand{\ig}{g}		\newcommand{\nlg}{g'}
		\newcommand{\tE}{E}				
						
	\newcommand{\tT}{T}						
							
					\newcommand{\id}{d}		\newcommand{\nd}{d'}
			\newcommand{\ico}{\clP}		\newcommand{\nco}{\clP'\!}
	\newcommand{\tdel}{\ul{\nabla}}

\newcommand{\adel}{\tilde{\tdel}}	\newcommand{\rdel}{\Tilde{\Tilde{\tdel}}}
\newcommand{\aT}{\tilde{\tT}}		\newcommand{\rT}{\Tilde{\Tilde{\tT}}}

\newcommand{\I}{I}

	\newcommand{\tm}{{\ul{m}}}

	\newcommand{\ta}{{\ul{a}}}	
	\newcommand{\tb}{{\ul{b}}}	
	\newcommand{\tc}{{\ul{c}}}	
	\newcommand{\td}{{\ul{d}}}	

	\newcommand{\ba}{{\bar{a}}}	
	\newcommand{\bb}{{\bar{b}}}	
	\newcommand{\bc}{{\bar{c}}}

\usepackage[all]{xy}

\begin{document}
%%%%%%%%%%%%%%%%%%%%%%%%%%%%%%%%%%
%
% Title
%
%%%%%%%%%%%%%%%%%%%%%%%%%%%%%%%%%%
\begin{titlepage}
\title{{\Huge \bf New insights in brane and Kaluza--Klein theory through almost product structures}}
\author{
        Magnus Holm\thanks{holm@fy.chalmers.se}
}
\date{  Chalmers University of Technology\\
        University of Gothenburg\\
        Dec 1998\vspace{1cm}\\
        \leavevmode
}
\maketitle
%%%%%%%%%%%%%%%%%%%%%%%%%%%%%%%%%%
%
% Abstract
%
%%%%%%%%%%%%%%%%%%%%%%%%%%%%%%%%%%
\begin{abstract}
We will show that gauge theory can be described by an almost product structure, which is a certain type of
endomorphism of the tangent bundle. We will recover the gauge field strength as the Nijenhuis tensor of
this endomorphism. We discuss a generalization to the case of a general Kaluza-Klein theory.
Furthermore, we will look at the classification of these almost product structures in the
case where we have a manifold with metric, and fit the M-brane solutions into this classification scheme.
In this analysis certain algebraic properties of the space of differential forms and multivectors are obtained.
All analysis is global but we will give local expressions where we find it suitable.
\end{abstract}
\end{titlepage}
%%%%%%%%%%%%%%%%%%%%%%%%%%%%%%%%%%%%%
%
% Section: Introduction
%
%%%%%%%%%%%%%%%%%%%%%%%%%%%%%%%%%%%%%
\section{Introduction}

In this paper we will look at brane theory, gauge and Kaluza--Klein theory from a new perspective. The basic idea is
that instead of looking at embedded branes or gauge theory over a base manifold, we will treat the total space directly.
As is well known, this is how Kaluza--Klein theory works. There, the total
manifold is constrained to have one part with a certain isometry group, that
becomes the gauge group upon compactification.

Here we will generalize this analysis,
in a completely global treatment, and show that Kaluza--Klein theory or normal gauge theory are nothing but
special cases of almost product manifolds. The characterization of Kaluza--Klein theory will be that we have one foliation
which is geodisable ({\it i.e.}, the almost product structure becomes a sort of Ehresmann connection), which serves as the fiber,
and one perhaps non-integrable distribution. These split the tangent bundle of the total space into two different parts, in fiber
bundle language called the vertical and horizontal respectively. The base space of a principal bundle is recovered as the
leaf space of the foliation, and the field strength as the Nijenhuis tensor of the almost product structure.
From this case of a geodisable foliation we will also find that imposing integrability on the normal distribution, the vanishing
of the Nijenhuis tensor gives us two new coboundary operators under which the entire graded algebra of differential forms will become
doubly graded. This gives us directly a topological splitting of the manifold into two parts, see also \cite{BlHe84}, why the
cohomology groups split under these two coboundary operators.

In brane theory we discuss how the solutions in fact may be regarded as foliations of the
total space rather than as embedded objects. In the case of {\it e.g.} the M5-branes we know that the solutions are non-singular \cite{DuKhLu95,Du96} but even in the cases
where the solutions are singular it is clear that although the objects may be introduced as sources by Dirac delta functions, consistency of the theory in the total space demands that we cut out these points or sections of the manifold. This again brings us back to foliated space solutions. In M-brane theory we will see that the discussed solutions are
indeed doubly foliated---the Nijenhuis tensor of the almost product structure, which will characterize the solution, vanishes.
In the case of brane solutions, though, we are also interested in the metric and we will thus put the metric solution into the classification regime
of almost product manifolds. We will also see that the solutions are characterized by certain basic forms, the anti-symmetric tensor fields,
which can be seen to be compatible with the almost product structure characterized by the brane solution or equivalently as
defining this almost product structure. In this new formalism we are also able to argue for the existence of new solutions to
M-theory in which we only require the brane to be integrable and the foliation needs not even be Ehresmann. In these
cases, ref. \cite{Co87} argues that in situations like this, when the foliation does not define a fibration of the manifold,
we could very well be up to a leaf space that is non-commutative.

The paper is divided into five section, of which you are now reading the first. When studying endomorphisms of the tangent bundle,
which is the base for these almost product structures that will be dealt with, we find a certain dual structure on the set of differential forms
and the set of multivectors on a manifold. In section two we will describe this dual picture and show how the set of derivations
on differential forms and multivectors are recovered in a very easy and similar way. These relations will then be of utmost interest
for us as we turn to the case of endomorphisms on the tangent bundle later in section four. But first we will review the basic concepts
of distributions and foliations on a manifold expressed in a global way. This occurs in section three where we also get acquainted
with the concept of basic and semi-basic forms. These are the keys of putting the anti-symmetric tensor fields into the context of foliations
and therefore the brane solutions. We will also see that the set of basic forms are closed under the exterior derivative and we thus
get the cohomology groups of the leafspace. This treatment in the first subsection of section three will be done completely without
the presence of a metric. In the following subsection, though, we will see what additional structure the presence of a metric will give us in terms of
distributions and foliations. Here we introduce the deformation tensor and look at the interpretation of its irreducible parts.

Section four is the core of the paper. Here we start in the first subsection to introduce endomorphisms on the tangent bundle in
a general framework. We introduce the I-bracket associated with an endomorphism and the Nijenhuis tensor, measuring how
far this endomorphism is from being a Lie algebra homomorphism on the infinite-dimensional Lie algebra of vector fields on
a manifold. We will also treat the case where a metric is present, in which we introduce the Jordan bracket associated with an endomorphism
and the Jordan tensor, measuring how much the Jordan bracket fails to commute with the endomorphism. Finally, we introduce a generalized
deformation tensor, which later will reduce to the deformation tensor introduced in section three. In subsection two we will
start to see how certain endomorphisms, namely almost product structures, will serve as characterizing possible foliations on a manifold.
We will here recover the tensors from section three, where we see that the Nijenhuis tensor measures non-integrability and the Jordan tensor
measures how far the two complementary distributions, defined by the almost product structure, are from being geodesic. We will
introduce two new connections which both commute with the almost product structure and which will be of certain interest in the classification
scheme in following subsection. In the last subsection, we present the classification and examine
certain important consequences of some of its special cases. We will see the splitting of the cohomology groups in the case where the almost product
structure defines two Ehresmann foliations, we will see how the holonomy groups split when the almost product structure is covariantly
constant, we will see how the brane solutions fit into this classification scheme and we will see that the Nijenhuis tensor indeed measures
the field strength in gauge theories and Kaluza--Klein theories. We will also present the local structures of the involved objects in some
selected cases.

In section five we will end by discussing how this new formalism can help us in understanding M-theory, %and deeper, 
and how one could proceed in studying these objects further.
%%%%%%%%%%%%%%%%%%%%%%%%%%%%%%%%%%%%%
%
% Section: Derivations on exterior forms
%
%%%%%%%%%%%%%%%%%%%%%%%%%%%%%%%%%%%%%
\section{Derivations on the exterior algebra of forms and vectors}

In this section we will take a look at the set of all derivations on both the set of differential forms and the set of multivectors on a manifold.
We will find two different types of brackets, namely the Schouten--Nijenhuis bracket, which is a bracket between multivectors, and the
Fr\"olicher--Nijenhuis bracket, which is a bracket between vector-valued forms. The Schouten--Nijenhuis bracket will by the adjoint mapping
become a derivation on the set of multivectors, while the Fr\"olicher--Nijenhuis bracket turns up in the commutator of two derivations on the set of 
differential forms. See \cite{Mic86,Mic87} for a more detailed study. The new thing here is that we will put the action on forms and multivectors
on an equal footing. We will see that all maps have its dual in the co-picture. But let us start with some preliminaries.
\defn{deriv}{Let $\ws$ be a manifold and let us denote
	\eqal{2}{
		\O^p:=&\O^p(\ws)\quad	&&\text{the $p$-forms on $\ws$}\\
		\O:=&\underset{p}{\oplus}\O^p\quad&&\text{the graded algebra of $p$-forms}\\
		\L^q:=&L^q(\ws)	&&\text{the $q$-vectors on $\ws$}\\
		\L:=&\underset{q}{\oplus}\L^q\quad&&\text{the graded algebra of $q$-vectors}
	}
}
We will now study maps on $\O$ and $\L$, especially those maps that are derivations. So we need some basic definitions.
\defn{gradedmaps}{Let $D\in{\rm Lin}(\O,\O),\;(D\in{\rm Lin}(\L,\L))$ be a linear map on the graded algebra of $p$forms ($q$-vectors).
	Then $D$ is said to be graded of degree $k$ if
	\eqnono{
		D:\O^p\longmapsto\O^{p+k},\quad (\L^p\longmapsto\L^{p+k}).
	}
	Let $D_i\in {\rm Lin}(\O,\O),\;(D_i\in {\rm Lin}(\L,\L))$ be graded linear maps of degree $k_i$, then we can define the graded commutator by
	\eqnono{
		[D_1,D_2]:=D_1\circ D_2-(\minus 1)^{k_1k_2}D_2\circ D_1,
	}
	which is again a graded linear map but of degree $k_1+k_2$. We can also define 
	the graded Jacobi bracket by
	\eqal{1}{
		[D_1,D_2,D_3]:=&[[D_1,D_2],D_3]+(\minus 1)^{k_1(k_2+k_3)}[[D_2,D_3],D_1]+\\
			&+(\minus 1)^{k_3(k_1+k_2)}[[D_3,D_1],D_2]
	}
	A graded linear map is said to be a graded derivation of degree $k$ if
	\eqnono{
		D(\o_1\wedge\o_2)=D\o_1\wedge\o_2+(\minus 1)^{kl}\o_1\wedge D\o_2,\qquad {\rm for}\; \o_1\in\O^l,(\L^l),\;\o_2\in\O,(\L).
	}
	We will denote the space of all derivations of degree $k$ by ${\rm Der}_k\O,\;({\rm Der}_k\L)$ and the space of all derivations by
	\eqnono{
		{\rm Der}\thinspace\O:=\underset{k}{\oplus}{\rm Der}_k\O,\quad({\rm Der}\thinspace\L:=\underset{k}{\oplus}{\rm Der}_k\L).
	}
}
Now it is easily seen that the set of all derivations on $\O$ and $\L$ respectively, forms a graded Lie algebra under the graded commutator.
\prop{liealgebraofder}{${\rm Der}\thinspace\O\;({\rm Der}\thinspace\L)$ becomes a graded Lie algebra with the graded commutator defined in \ref{gradedmaps}.
	This means that it satisfies the graded Jacobi identity, {\it i.e.},
	\eqnono{
		[D_1,D_2,D_3]=0,\quad\forall D_i\in{\rm Der}\thinspace\O,\;(\forall D_i\in{\rm Der}\thinspace\L).
	}
}
\proof{By direct calculation.}
We will study these sets of derivations in the next two subsections. We will see that we can introduce a map called the generalized Lie derivative
which is not necessarily a derivation but has some nice characteristics. Among these are the natural fact that it reduces to the usual Lie derivative
in the case it acts by a vector, and the fact that it has a dual map. All this will become clear in subsection 2.
%%%%%%%%%%%%%%%%%%%%%%%%%%%%%%%%%%%%%%
%
% Subsection: Derivations on Lambda
%
%%%%%%%%%%%%%%%%%%%%%%%%%%%%%%%%%%%%%%
\subsection{Derivations on $\L$}

Here we will see how we can obtain the Schouten--Nijenhuis bracket of two multivectors from the generalized Lie derivative
to be introduced.
We will start by introducing a formal boundary operator on the set of multivectors on $\ws$ denoted by $\L$.
\defn{deriva}{Let $\L$ be the graded algebra of all $p$-vectors on $\ws$. Then we can formally form a differential complex
	over the vector fields with the sequence
	\eqnono{\begin{CD}
		0@<\pa<<\L^1@<\pa<<\L^2\cdots@<\pa<<\L^q@<\pa<<\cdots
	\end{CD}}
	where $\pa$ is a ``boundary'' operator with the characteristics
	\eqnono{
		\pa:\L\longmapsto\L,\quad\L^q\longmapsto\L^{q-1},\quad\pa\circ\pa=0.
	}
	It is defined on a $p$-vector by
	\eqnono{
		\pa(X_1\wedge\ldots\wedge X_p):=\sum_{i<j}(\minus 1)^{i+j+1}[X_i,X_j]\wedge X_1\wedge\no{i}\no{j}\wedge X_p,
	}
	where $\no{i}\no{j}$ means that $X_i$ and $X_j$ are omitted, and satisfies $\pa\L^1=0$. The nilpotency follows from the
	Jacobi identity of the vector bracket.
}
\remk{rembound}{It should be pointed out that $\pa$ defined above is no derivation.
	It is not defined on functions and not even well defined on general $p$-vectors.}
We refer to the definition of $\pa$ as formal because of what we learned from remark \ref{rembound}. We can see why it is not well defined on $p$-vectors by taking
the 2-vector example. Let $X_1\wedge X_2\in\L^2$ be a 2-vector on $\ws$, then we know that as a 2-vector $X_1\wedge X_2=fX_1\wedge f^{\minus 1}X_2$,
where $f$ is an arbitrary function on $\ws$. But $\pa (X_1\wedge X_2)=[X_1,X_2]\neq \pa (fX_1\wedge f^{\minus 1}X_2)=[X_1,X_2]+f^{\minus 1}X_1[f]X_2
-fX_2[f^{\minus 1}]X_1$, so we see that it is not well defined. It should be noted however that it is well defined on the set of 
multivectors on a Lie algebra where 
the functions instead becomes pure numbers. Nevertheless we will see that the formal boundary operator is of importance. 

We also need
to introduce the exterior product between multivectors.
\defn{exter}{Let $\L$ be the graded algebra of all $p$-vectors on $\ws$, let $X\in\L^1$ be a vector, then the {\bf exterior product}
	with respect to $X$ have the following characteristics:
	\eqnono{
		\ve_X:\L\longmapsto\L,\quad\L^q\longmapsto\L^{q+1},\quad\ve_X\circ\ve_X=0.
	}
	The exterior product is defined by its action on a $p$-vector by
	\eqnono{
		\ve_X(X_1\wedge\ldots\wedge X_p):=X\wedge X_1\wedge\ldots \wedge X_p.
	}
	Let $Y\in\L^1$ be another vector, then
	\eqnono{
		[\ve_X,\ve_Y]=0
	}
	{\it i.e.}, $\ve_X\ve_Y=-\ve_Y\ve_X$. Now let $X_i\in\L^1$ be $p$ vectors and let us extend the exterior product in the sense
	\eqnono{
		\ve_{X_1\wedge\ldots\wedge X_p}:=\ve_{X_1} \circ \ldots \circ \ve_{X_p}.
	}
	This makes the exterior product a $p$-graded map, {\it i.e.},
	\eqnono{	
		\ve_{X_1\wedge\ldots\wedge X_p}:\L\longmapsto\L,\quad\L^q\longmapsto\L^{q+p}.
	}
}
\remk{remext}{It should be noted that the map $\ve_X$, although a linear map, is no derivation.}
We can now create the generalized Lie derivative by taking the commutator of these two maps on $\L$. 
\defn{vectorlie}{Let $X\in\L$ be a $p$-vector on $\ws$. We can then define the generalized Lie derivative, $\vlie_X$, with following characteristics:
	\eqnono{
		\vlie_X:\L\longmapsto\L,\quad\L^q\longmapsto\L^{q+p-1}.
	}
	It is defined simply through the boundary operator and the exterior product by
	\eqnono{
		\vlie_X:=[\pa,\ve_X].
	}
}
\remk{remlie}{The generalized Lie derivative, $\vlie_X$ is only a derivation in the case when $X\in\L^1$ is a vector. In this case
	it is of course the usual Lie derivative.}
So to sum up we have three maps on $\L$, the boundary operator $\pa$, the exterior product $\ve_X$ and the generalized Lie derivative $\vlie_X$, of whom neither
is a derivation except for the case when $X$ is a pure vector when the generalized Lie derivative reduces to the ordinary Lie derivative. Now, however, if
we take the commutator of two generalized Lie derivatives we recover the Schouten--Nijenhuis bracket which as the adjoint mapping is a derivation on
$\L$.
\defn{schnij}{Let $X\in\L^p,\;Y\in\L^q$ be two multivectors on $\ws$ and let $\o\in\O^{p+q-1}$ a closed $(p+q-1)$-form, then we
	can define the {\bf Schouten--Nijenhuis bracket} with following characteristics:
	\eqnono{
		[X,Y]:\L\times\L\longmapsto\L,\quad\L^p\times\L^q\longmapsto\L^{p+q-1},
	}
	or in the sense of adjoint mapping
	\eqal{1}{
		{\rm ad}_X:\L\longmapsto\L,&\quad\L^q\longmapsto\L^{q+p-1}\\
			{\rm ad}_XY&:=[X,Y].
	}
	The following definitions of the Schouten--Nijenhuis bracket are equivalent
	\eqal{2}{
		(i)&&\vlie_{[X,Y]}&:=[\vlie_X,\vlie_Y]\\
		(ii)&&\qquad(\minus 1)^{p-1}[X,Y]&:=\pa(X\wedge Y)-\pa X\wedge Y-(\minus 1)^{p}X\wedge \pa Y\\
		(iii)&&[X,Y]&:=\sum_{i,j}(\minus 1)^{i+j}[X_i,Y_j]\wedge X_1\wedge\no{i}\wedge X_p\wedge Y_1\wedge\no{j}\wedge Y_q\\
		%(iv)&&i_{[X,Y]}\o&:=(i_Xdi_Y-(\minus 1)^?i_Ydi_X)\o\\
	}
}
\prop{schnijen}{The Schouten--Nijenhuis bracket as a map ${\rm ad}_X:\L\mapsto\L$ is a derivation on $\L$ and satisfies the graded Jacobi identity. 
	It forms thus a Lie algebra structure on ${\rm Der}\thinspace\L$. 
}
We will see in next subsection, where we look at derivations on the set of differential forms $\O$ on $\ws$, that the boundary operator will become a kind
of dual to the exterior derivative or co-boundary operator on $\O$, the exterior product will be dual to the interior product and the generalized Lie
derivative will become dual to a generalized Lie derivative on differential forms. 
%%%%%%%%%%%%%%%%%%%%%%%%%%%%
%
% Subsection: The dual maps of \L and \O
%
%%%%%%%%%%%%%%%%%%%%%%%%%%%%
\subsection{The dual maps on $\L$ and $\O$}

We will here define
the exterior derivative, the interior product and the generalized Lie derivatives as dual maps to those defined in the previous subsection.
\defn{cobound}{Let $\O$ be the graded algebra of all $p$-forms on $\ws$ and let us form the differential complex over $\O$ with sequence
 	\eqnono{\begin{CD}
		0@>i_*>>\O^0@>d>>\O^1@>d>>\cdots\O^p@>d>>\cdots
	\end{CD}
	}
	where $i_*$ is an inclusion, and $d$ is the coboundary operator on $\O$ with following characteristics:
	\eqnono{
		d:\O\longmapsto\O,\quad\O^p\longmapsto\O^{p+1},\quad d\circ d=0.
	}
	Let $X_i\in\L^1$ be vector fields on $\ws$ then we can define the coboundary operator by
	\eqsnono{
		d\o(X_1,\ldots,X_{p+1})&:=&\sum_i(\minus 1)^{i+1}\Lie_{X_i}\o(X_1,\no{i},X_{p+1})-\\
					&&-\o(\pa(X_1\wedge\ldots\wedge X_{p+1}))\\.
	}
	The nilpotency is not manifest but follows by the relation $[\Lie_{X_i},\Lie_{X_j}]=\Lie_{[X_i,X_j]}$. We see that the
	coboundary operator in some sense is the adjoint operator of $\pa$.
}
We see that although the boundary operator defined in \ref{deriva} was not well defined, the total expression for the exterior derivative is.
The exterior derivative is of course a derivation, hence its name. Now to the interior product.
\defn{intprod}{Let $\O$ be the graded algebra of $p$-forms on $\ws$, $X\in\L^1$ be a vector field and let $\ve_X$ be the exterior
	product defined in \ref{exter}. Then we can define the interior product with the following characteristics:
	\eqnono{
		i_X:\O\longmapsto\O,\quad\O^p\longmapsto\O^{p-1},\quad i_X\circ i_X=0,
	}
	as the adjoint of the exterior product, {\it i.e.},
	\eqnono{
		i_X\o(X_1,\ldots,X_{p-1})=\o(\ve_X(X_1\wedge\ldots\wedge X_{p-1})).
	}
	The interior product satisfies $i_X\O^0=0$ and
	\eqnono{
		[i_X,i_Y]=0,
	}
	where $Y\in\L^1$. We can in fact extend the interior product in the same way as we extended the exterior product so that for $X_i\in\L^1$
	\eqnono{
		i_{X_1\wedge\ldots\wedge X_q}:=i_{X_1} \circ \ldots \circ i_{X_q}
	}
	satisfies
	\eqnono{
		i_{X_1\wedge\ldots\wedge X_q}\o(X_{q+1},\ldots,X_{p})=\o(\ve_{X_1\wedge\ldots\wedge X_q}X_{q+1}\wedge\ldots\wedge X_{p})
	}
	and is therefore a $q$ graded map, {\it i.e.},
	\eqnono{
		i_{X_1\wedge\ldots\wedge X_q}:\O\longmapsto\O,\quad\O^p\longmapsto\O^{p-q}.
	}
}
\remk{remint}{The interior product, $i_X$, is a derivation only when $X\in\L^1$ is a vector field.}
If we recall the definition of the ordinary Lie derivative acting on forms we immediately see how this can be generalized. 
\defn{liederiv}{Let $\O$ be the graded algebra of $p$-forms on $\ws$, let $X\in\L^1$ be a vector field. Let $\Lie_X$ be the
	Lie derivative with following characteristics:
	\eqnono{
		\Lie_X:\O\longmapsto\O,\quad\O^p\longmapsto\O^{p}.
	}
	It is defined by
	\eqnono{
		(\Lie_X\o)(X_1,\ldots,X_{p})=\Lie_X\o(X_1,\ldots,X_p)-\o(\Lie_X(X_1\wedge\ldots\wedge X_p))
	}
	and satisfies Cartan's infinitesimal homotopy formula
	\eqnono{
		\Lie_X=[i_X,d].
	}
	Let $Y\in\L^1$ be another vector field. The Lie derivative satisfies the following equations:
	\eqsnono{
		[\Lie_X,d]&=&0\cr
		[\Lie_X,i_Y]&=&i_{[X,Y]}\cr
		[\Lie_X,\Lie_Y]&=&\Lie_{[X,Y]}
	}
}
So we can proceed, as in the previous subsection, by introducing the generalized Lie derivative acting on forms by simply generalizing Cartan's formula. 
\defn{generallie}{Let $X\in\L^q$ be a multivector on $\ws$, then the generalized Lie derivative on $p$-forms is a map
	with following characteristics:
	\eqnono{
		\flie_X:\O \longmapsto \O,\quad \O^p \longmapsto \O^{p-q+1},
	}
	and is defined by
	\eqnono{
		\flie_X:=[i_X,d].
	}
}
\remk{genabove}{This generalized Lie derivative, $\flie_X$, acting on forms is only a derivation on $\O$ when $X$
	is a vector field.}
By this remark we have a similar case to that of the previous subsection, now however we do not know for sure that this map is dual to the
generalized Lie derivative acting on multivectors introduced before, but we have to show this.
\prop{genlierel}{Let $X=X_1\wedge\ldots \wedge X_q\in\L^q$ be a $q$-vector and let $\o\in\O^p$ be a $p$-form then the 
	generalized Lie derivative satisfies
	\eqal{1}{
		(\flie_X\o)(X_{q+1},\ldots,X_{p+1})=&\sum_{i=1}^q(\minus 1)^{i+1}\Lie_{X_i}\o(X_1,\no{i},X_q,X_{q+1},\ldots,X_{p+1})-\\
						&-\o(\vlie_X(X_q\wedge\ldots\wedge X_{p+1}))
	}
}
\proof{By direct calculation:
	\eqal{1}{
  \!\!\!\!\!\!(\flie_X\o)&(X_{q+1},\ldots,X_{p+1})=(i_{X_1\wedge\ldots \wedge X_q}d-(\minus 1)^qdi_{X_1\wedge\ldots \wedge X_q})(\o)(X_{q+1},\ldots,X_{p+1})=\\
				=&d\o(\ve_{X_1\wedge\ldots \wedge X_q}X_{q+1}\wedge\ldots\wedge X_{p+1})-\\
				&(\minus 1)^q\sum_{i=q+1}^{p+1}(\minus 1)^{i-q+1}\Lie_{X_i}(i_{X_1\wedge\ldots \wedge X_q}\o)(X_{q+1},\no{i},X_{p+1})+\\
				&(\minus 1)^q(i_{X_1\wedge\ldots \wedge X_q}\o)(\pa(X_{q+1}\wedge\ldots\wedge X_{p+1}))=\\
		=&\sum_{i=1}^{p+1}(\minus 1)^{i+1}\Lie_{X_i}\o(X_1,\no{i},X_{p+1})-\o(\pa\ve_{X_1\wedge\ldots \wedge X_q}X_{q+1}\wedge\ldots\wedge X_{p+1})-\\
			&-\sum_{i=q+1}^{p+1}(\minus 1)^{i+1}\Lie_{X_i}\o(X_1,\ldots,X_q,X_{q+1},\no{i},X_{p+1})+\\
			&(\minus 1)^q\o(\ve_{X_1\wedge\ldots \wedge X_q}\pa X_{q+1}\wedge\ldots\wedge X_{p+1})=\\
			=&\sum_{i=1}^q(\minus 1)^{i+1}\Lie_{X_i}\o(X_1,\no{i},X_q,X_{q+1},\ldots,X_{p+1})-\\
						&-\o(\vlie_X(X_q\wedge\ldots\wedge X_{p+1}))
	}
}
Put together we have now seen that all these maps come with their duals. This is pointed out in following remark.
\remk{boundcobound}{We see that these operators are formally adjoints to each others as acting on forms and multivectors respectively
	and we can write
	\eqal{2}{
		\O&&&\L\\
		d&&\quad\longleftrightarrow\quad&\pa\\
		i_X&&\quad\longleftrightarrow\quad&\ve_X\\
		\hat{\Lie}_X:=[i_X,d]&&\quad\longleftrightarrow\quad&[\pa,\ve_X]=:\check{\Lie}_X
	}
	as a correspondence table.
}
Now as we saw that we recovered the Schouten--Nijenhuis bracket when taking the commutator of two generalized Lie derivatives acting on multivectors
we shall find out that we will get the same thing for the generalized Lie derivative acting on forms (up to a sign).
\prop{firstcomutatorrel}{Let $X\in\L^p$ and $Y\in\L^q$ be two multivectors. Then the brackets defined through the generalized Lie derivatives, {\it i.e.},
	\eqal{1}{
		\vlie_{[X,Y]\check{}}&:=[\vlie_X,\vlie_Y]\\
		\flie_{[X,Y]\hat{}}&:=[\flie_X,\vlie_Y]
	}
	are related by
	\eqnono{
		[X,Y]\hat{}=-[Y,X]\check{}=(\minus 1)^{(p-1)(q-1)}[X,Y]\check{}
	}
} 
\proof{By combinatorics.}
%%%%%%%%%%%%%%%%%%%%%%%%%%%%%%%%%%
%
% Subsection: Derivations on Omega
%
%%%%%%%%%%%%%%%%%%%%%%%%%%%%%%%%%%
\subsection{Derivations on $\O$}

In this subsection we will look at the set of all derivations on the set of differential forms $\O$ on $\ws$. 
We will see that they are spanned by mappings involving vector valued forms denoted $\O^p_1$, but as before we will start by looking at these
mappings acting on $\L$ and then see that their duals acting on forms are derivations. So lets first start with the exterior product.
\defn{exi}{Let $I\in\O^p_1$ be a vector-valued $p$-form on $\ws$, then the exterior product $\ve_I$ of $I$ is a map with following characteristics:
	\eqnono{
		\ve_I:\L\longmapsto\L,\quad\L^q\longmapsto\L^{q-p+1},
	}
	and if  $\m\in {\rm Perm}(p+q)$ we can define the exterior product of $I$ on a $(p+q)$-vector by
	\eqnono{
		\ve_I(X_1\wedge\ldots\wedge X_{p+q}):={1\fr p!q!}\sum_\m(\minus 1)^\m I(X_{\m_1},\ldots,X_{\m_{p}})\wedge\ldots\wedge X_{\m_{p+q}}
	} 
}
\remk{exirem}{If $I\in\O^1_1$ is a endomorphism, {\it i.e.}, a 1-1 tensor, then $\ve_I$ is a derivation on $\L$.}
We can now define the generalized Lie derivative of a vector-valued form acting on differential forms by the immediate analogue of definition \ref{generallie}.
\defn{lie}{Let $I\in\O^p_1$ be a vector-valued $p$-form on $\ws$, then let us define the generalized Lie derivative acting as
	a map on $\L$ with following characteristics:
	\eqnono{
		\vlie_{I}:\L\longmapsto\L,\quad\L^q\longmapsto\L^{q-p}
	}
	We define it in analogous way as before by
	\eqnono{
		\vlie_{I}:=[\pa,\ve_I]
	}
}
From the definition above we can now find the expression for the generalized Lie derivative in terms of the ordinary commutator on vectors.
\prop{genlie}{Let $I\in\O^p_1$ be a vector-valued $p$-form on $\ws$, let $X_i\in\L^1$ be vector fields and $\m\in {\rm Perm}(p+q)$. Then
	\eqal{1}{
		\vlie_I(X_1\wedge\ldots&\wedge X_{p+q})={1\fr p!(q-1)!}\sum_\m(\minus 1)^\m[I(X_{\m_1},\ldots,X_{\m_p}),X_{\m_{p+1}}]
			\wedge\ldots\wedge X_{\m_{p+q}}-\\
		&-{(\minus 1)^{p-1}\fr (p-1)!(q-1)!2!}\sum_\m(\minus 1)^\m I([X_{\m_1},X_{\m_2}],\ldots,X_{\m_{p+1}})
			\wedge\ldots\wedge X_{\m_{p+q}}
	}
}
\proof{\cite{Mic86} plus the proof of \ref{lielie}}
Now we are ready to study the set of derivations on $\O$. We will not go into details but only stress the differences appearing with this
dual picture and refer to \cite{Mic86} for a more detailed study. To start with we will define what we mean by a algebraic derivation.
\defn{alg}{Let $D\in{\rm Der}\thinspace\O$ then $D$ is said to be algebraic if
	\eqnono{
		D|\O^0=0
	}
	Let $\o\in\O$ be a $p$-form on $\ws$, then if $D$ is {\bf algebraic} we have
	\eqnono{
		D(f\o)=fD\o,\quad\forall f\in C^\infty(\ws)
	}
	which means that $D$ is tensorial.
}
We will see that the set of algebraic derivations on $\O$ is spanned by the interior product of vector valued forms on $\ws$, so let us
define the interior product again as the dual map to the exterior product.
\defn{int}{Let $I\in\O^p_1$ be a vector-valued $p$-form on $\ws$ and let $\o\in\O^q$ be a $q$-form. Then define the interior product of $I$ on
	$\O$ as a map with following characteristics:
	\eqnono{
		i_I:\O\longmapsto\O,\quad\O^q\longmapsto\O^{q+p-1}.
	}
	Let $\o\in\O^q$ be a $q$-form and define the internal product as the formal adjoint to the exterior product as
	\eqnono{
		i_I\o(X_1,\ldots,X_{p+q-1}):=\o(\ve_I(X_1\wedge\ldots\wedge X_{p+q-1})).
	}
}
Now \cite{Mic86} tells us that not only is the map $i_I$ an algebraic derivation on $\O$, but that all algebraic derivations can be written
in that way, so we have a one-to-one correspondence.
\prop{algder}{Let $D\in{\rm Der}_k\thinspace\O$ be a graded derivation of degree $k$, then
	\eqnono{
		D=i_I
	}
	for some $I\in\O^{k+1}_1$.
}
\proof{See \cite{Mic86}.}
It is also clear that if we again introduce the generalized Lie derivative of vector-valued forms by the analogue to Cartan's formula we know that
it must be a derivation because it is now a commutator of two derivations.
\defn{lieform}{Let $I\in\O^p_1$ be a vector-valued $p$-form on $\ws$ and define the generalized Lie derivative on $q$-forms as a map with following
	characteristics:
	\eqnono{
		\flie_I:\O\longmapsto\O,\quad\O^q\longmapsto\O^{q+p},
	}
	defined by
	\eqnono{
		\flie_I:=[i_I,d]
	}
}
What is not clear is that it again is dual to the generalized Lie derivative acting on multivectors defined in \ref{lie} which indeed is no derivation
on $\L$ unless $I$ is a vector.
\prop{lielie}{Let $I\in\O^p_1$ be a vector-valued $p$-form on $\ws$, $\o\in\O^q$ a $q$-form and $X_i\in\L^1$ be vectors. Then
	\eqal{1}{
		(\flie_I\o)(X_1,\ldots,X_{p+q})=&{1\fr p!q!}\sum_\m(\minus 1)^\m\Lie_{I(X_{\m_1},\ldots,X_{\m_p})}\o(X_{\m_{p+1}},\ldots,X_{\m_{p+q}})-\\
				&-\o(\vlie_I(X_1\wedge\ldots\wedge X_{p+q}))
	}
}
\proof{The proof is by direct calculation,
	\eqal{1}{
		(\flie_I\o)&(X_1,\ldots,X_{p+q})=((i_Id-(\minus 1)^{p-1}di_I)\o)(X_1,\ldots,X_{p+q})=\\
			=&d\o(\ve_I(X_1\wedge\ldots \wedge X_{p+q}))\\
			&-(\minus 1)^{p-1}\left(\sum_i(\minus 1)^{i+1}\Lie_{X_i}(i_I\o)(X_1,\no{i},X_{p+q})
			-i_I\o(\pa(X_1\wedge\ldots\wedge X_{p+q}))\right)=\\
			=&{1\fr p!q!}\sum_\m(\minus 1)^\m d\o(I(X_{\m_1},\ldots,X_{\m_{p}})\wedge\ldots\wedge X_{\m_{p+q}})\\
			&-(\minus 1)^{p-1}\left(\sum_i(\minus 1)^{i+1}\Lie_{X_i}(i_I\o)(X_1,\no{i},X_{p+q})
			-\o(\ve_I\pa(X_1\wedge\ldots\wedge X_{p+q}))\right)=\\
			=&{1\fr p!q!}\sum_\m(\minus 1)^\m \left(\Lie_{I(X_{\m_1},\ldots,X_{\m_{p}})}\o(X_{\m_{p+1}}\wedge\ldots\wedge X_{\m_{p+q}})+\right.\\
			&\left.+q(\minus 1)^{p-1}(\minus 1)^{\m_i}\Lie_{X_{\m_{i}}}\o(I(X_{\m_1},\ldots,X_{\m_{p}})
				\wedge\no{\m_i}\wedge X_{\m_{p+q}})\right)\\
			&-\o(\pa\ve_I(X_1\wedge\ldots \wedge X_{p+q}))\\
			&-(\minus 1)^{p-1}\left(\sum_i(\minus 1)^{i+1}\Lie_{X_i}\o(\ve_I(X_1,\no{i},X_{p+q}))
			-\o(\ve_I\pa(X_1\wedge\ldots\wedge X_{p+q}))\right)=\\
			=&{1\fr p!q!}\sum_\m(\minus 1)^\m\Lie_{I(X_{\m_1},\ldots,X_{\m_p})}\o(X_{\m_{p+1}},\ldots,X_{\m_{p+q}})-\\
			&-\o(\vlie_I(X_1\wedge\ldots\wedge X_{p+q}))
	}
}
From \cite{Mic86} we also know that any derivation can be split into two parts, one part which is algebraic and one which looks like the generalized
Lie derivative.
\prop{derivs}{Let $D\in{\rm Der}_k\O$ be a derivation of degree $K$ on $\ws$ then it can be uniquely be decomposed like
	\eqnono{
		D=\flie_I+i_J
	}
	for some $I\in\O^k_1$, $J\in\O^{k+1}_1$. Furthermore we have the following equivalences
	\eqsnono{
		I=0&\Longleftrightarrow&D\;algebraic\\
		J=0&\Longleftrightarrow&[D,d]=0
	}
}
\proof{See \cite{Mic86}.}
Again we can introduce a bracket by looking at the commutator of two generalized Lie derivations. This bracket is the Fr\"olicher--Nijenhuis bracket.
\defn{fronij}{Let $I\in\O^p_1,\;J\in\O^q_1$ be two vector-valued forms on $\ws$ and let $\O_1$ denote the set of all vector-valued forms
	on $\ws$ then we define
	the {\bf Fr\"olicher--Nijenhuis bracket} with following characteristics:
	\eqnono{
		[I,J]:\O_1\times\O_1\longmapsto\O_1,\quad\O^p_1\times\O^q_1\longmapsto\O^{p+q}_1
	}
	Let $\m\in {\rm Perm}(p+q)$, then the following 
	definitions of the Fr\"olicher--Nijenhuis bracket are equivalent
	\eqal{2}{
		(i)&&\flie_{[I,J]}:=&[\flie_I,\flie_J]\\
		(ii)&&\qquad[I,J](X_1,\ldots,X_{p+q}):=&{1\fr p!q!}\sum_\m(\minus 1)^\m[I(X_{\m_1},\ldots,X_{\m_p}),J(X_{\m_{p+1}},\ldots,X_{\m_{p+q}})]+\\
			&&&-J(\vlie_I(X_1\wedge\ldots\wedge X_{p+q}))+(\minus 1)^{pq}I(\vlie_J(X_1\wedge\ldots\wedge X_{p+q}))
	}
}
\proof{\cite{Mic86} plus \ref{genlie}}
We will also find that if we define the bracket by the commutator of two generalized Lie derivations acting on multivectors we will again get the
Fr\"olicher--Nijenhuis bracket up to a sign.
\prop{comutatorrel}{Let $I\in\O^p_1$ and $J\in\O^q_1$ be two vector-valued forms then the brackets defined through the generalized Lie derivatives, {\it i.e.},
	\eqal{1}{
		\vlie_{[I,J]\check{}}&:=[\vlie_I,\vlie_J]\\
		\flie_{[I,J]\hat{}}&:=[\flie_I,\vlie_J]
	}
	are related by
	\eqnono{
		[I,J]\hat{}=-[J,I]\check{}=(\minus 1)^{pq}[I,J]\check{}
	}
}
\proof{By combinatorics.}
%%%%%%%%%%%%%%%%%%%%%%%%%%%%%%%%%%%%%%%
%
% Section: Distributions and foliations
%
%%%%%%%%%%%%%%%%%%%%%%%%%%%%%%%%%%%%%%%
\section{Distributions and foliations}

In this section we will review the basic concepts of distributions and foliations on a manifold. We will start with a general treatment
in the first part, where neither a metric nor a connection is needed. We will see how we, from the solutions of the M2,5-branes, find that the anti-symmetric tensor field of the solution
itself defines a foliation on the manifold. In the latter part we will see how we can give these concepts more structure by 
adding a metric and a connection.

%%%%%%%%%%%%%%%%%%%%%%%%
%
% Subsection: General treatment
%
%%%%%%%%%%%%%%%%%%%%%%%%
\subsection{General treatment}

First we need to understand the basic concepts of distributions and foliations, so let us start by defining these.
\defn{distribution}{Let $\ws$ be a manifold with tangent bundle $T\ws$, then a {\bf distribution} on $\ws$ is a subset
	of the tangent bundle such that, for any point $x$ in $\ws$, the fiber $\esD_x = \esD \cap T_x\ws$ is a
	vector subspace of $T_x\ws$. The dimension of $\esD_x$ is called the {\bf rank} of the distribution. We
	will denote the distributions of constant rank $k$-distributions, where $k$ is the rank. 
}
\defn{foliation}{Let $\ws$ be a manifold with dimension $m$, then a ($k$-) foliation, $\esF$, is a family of connected subsets, $\esF=\{\clL_\a\}$,
	called leaves, such that
	\rnum{
		\item	$\underset{\a}{\cup}\clL_\a=\ws$
		\item	$\clL_\a\cap\clL_\b=0,\quad \a\neq\b$
		\item	For any point $x\in\ws$ there exists a local coordinate system (chart $(U_x,\vf)$)in which the leaves
			are coordinate surfaces.
	}
}
\begin{figure}
\begin{center}
        \includegraphics[width=11cm]{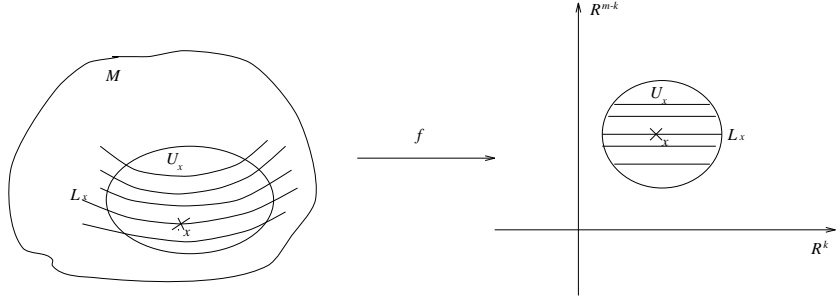} 
\end{center}
\caption{Foliation}
\end{figure}
It is clear from the definition of a foliation that it trivially defines a $k$-distribution and that this always locally can be spanned by coordinate
vectors. If $\ws$ is a manifold and $x^\tm$ are local coordinates in a patch $U$ then we will split it to $x^\tm=(x^m,y^{m'})$ where the leaves of
the foliation are determined by local coordinate surfaces like $y^{m'}=C^{m'}$. The distribution associated with the leaves is then spanned
by $\{\pa/\pa_m\}$ which are the annihilators of the normal pfaffian forms of the surfaces, $dy^{m'}$. We will see that this distribution is trivially 
integrable, but let us first define the concept.
\defn{integrability}{Let $\esD$ be a $k$-distribution on a manifold $\ws$, then the set of all vectors in $\esD$ forms a graded algebra
	on $\ws$ with the usual wedge product. We will denote this algebra
	\eqal{1}{
		\L_\esD^q&:=\L^q(\ws)|_\esD\\
		\L_\esD&:=\underset{q}{\oplus}\L_\esD^q
	}
	where $\L^q_\esD$ is the set of $q$-vectors lying in $\wedge^q\esD$. This algebra is a subalgebra of $\L$, {\it i.e.}, $\L_\esD \subset \L$.
	The distribution, $\esD$, is said to be {\bf integrable} if the algebra $\L_\esD$ is closed under the Schouten--Nijenhuis bracket,
	that is
	\eqnono{
		[\L_\esD,\L_\esD]\subset \L_\esD.
	}
}
\remk{reminteger}{The usual definition of integrability of a distribution is that, taken any two vectors $X,Y\in\L^1_\esD$,
	the commutator of these vector fields will still be a vector field of the distribution, or 
	\eqnono{
		[X,Y]\subset \L^1_\esD,\quad \forall X,Y\in\L^1_\esD,
	}
	but from the definition of the Schouten--Nijenhuis bracket we trivially see that the above definition of integrability
	is the same. The basic property of integrability is of course the existence of an integral manifold at every point, $x\in\ws$.
	Integrability also assures that this integral manifold is unique and that the dimension is equal to the rank of the integrable
	distribution.
}
Now obviously the distribution associated with the leaves of the foliation is integrable because it is locally spanned by coordinate vectors and
any two vectors built from these will be closed under the bracket in the sense that the resulting vector will again lie in the span of these
coordinate vectors. Now one can go even further and prove that in fact any distribution of constant rank that is integrable also defines a foliation.
\prop{distfoli}{Let $\esD$ be a $k$-distribution on a manifold $\ws$, then $\esD$ defines a foliation if and only if $\esD$ is
	integrable. Furthermore the leaves of this foliation are the integral manifolds of the distribution $\esD$.
}
\proof{See \cite{Ch46}.}
So we get a 1-1 relation between the concept of an integrable distribution of constant rank and a foliation.
\remk{vectorstructure}{We see from the definition of a foliation and the equivalence to an integrable distribution that if
	$\esF$ is an integrable distribution and $X\in\L_\esF$ is a vector field lying in the distribution then in every
	patch there exist coordinates $(x^m,y^{m'})$ such that the vector field $X$ can be expressed locally as
	\eqnono{
		X=X^{m}\pa_m=X^{m}(x,y){\pa\fr \pa x^m}.
	}
	The coordinate surfaces $y^{m'}=C^{m'}$ are the leaves and $\pa/\pa x^m$ are the basis vectors along the leaves.
}
We have seen that a $k$-distribution can be imposed as a subset of the set of $p$-vectors on $\ws$, which of course truncates at $k+1$,
and that it in fact is a subalgebra under the Schouten--Nijenhuis bracket if and only if it defines a foliation. But now we want to see how
we can understand this in the co-picture, where we look at the set of $p$-forms instead. So lets start with some basic definitions.
\defn{annihilator}{Let $\esD$ be a $k$-distribution on a manifold $\ws$, then the {\bf annihilator} or the 
	{\bf codistribution} of a distribution is denoted by, ${\esD^*}'$, and defined by
	\eqnono{
		{\esD^*}':=\bigcup_{x\in\ws}{\esD^*_x}'
	}
	where
	\eqnono{
		{\esD^*_x}':=\{\o\in T^*_x\ws :\; i_X\o=0,\; \forall X\in\L^1_\esD\}.
	}
	The set of all pfaffian forms in ${\esD^*}'$ forms a graded algebra on $\ws$ under the wedge product. The algebra is denoted by
	\eqal{1}{
		\O_{{\esD^*}'}^p&:=\O^p(\ws)|_{{\esD^*}'},\\
		\O_{{\esD^*}'}&:=\underset{p}{\oplus}\O^p_{{\esD^*}'}
	}
	where $\O^p_{{\esD^*}'}$ is the set of $p$-forms lying in $\wedge^p{\esD^*}'$. This algebra is a subset of the algebra of 
	differential forms on $\ws$, {\it i.e.}, $\O_{{\esD^*}'}\subset \O$.
}
\defn{ideal}{Let $\esD$ be a $k$-distribution on a manifold $\ws$, then the ideal of $\esD$ is defined by
	\eqal{1}{
		I_\esD:=&\underset{p}{\oplus}I^p_\esD,\\
		I^p_\esD=&\{\o\in\O^p:\o( X_1,\ldots, X_p)=0,\;\forall X_i\in\L^1_\esD\},
	}
	then $I_\esD\subset\O$ is to a subset of $\O$.
}
\remk{codist-ideal}{We see that $\O_{{\esD^*}'}\subset I_\esD$, so the ideal of $\esD$ is bigger than the set of forms spanned by the codistribution.
	We can picture these two types of forms in the case when $\esD$ is integrable and the annihilator locally is spanned by the pfaffian forms
	$\{dy^{m'}\}$, by 
	\eqal{2}{
		\o=&\o_{m_1'-m_p'} dy^{m_1'} \wedge \ldots \wedge dy^{m_p'}\\
		\eta=&\eta_{m_1'-m_p'} \vf^{m_1'}\wedge \ldots \wedge \vf^{m_{p-1}'} \wedge dy^{m_p'}
	}
	where $\o \in \O^p_{{\esD^*}'}$, $\eta \in I_\esD^p$ and $\vf^{m'}$ are arbitrary pfaffian forms.
}
The reason for introducing both these two types of subsets of the graded algebra of exterior forms on $\ws$ is that, although the subset
$\O_{{\esD^*}'}$ seems more natural, we need the ideal to test the integrability of the distribution. In fact we have following proposition.
\prop{cointeger}{Let $\esD$ be a $k$-distribution on a manifold $\ws$, let $I_\esD$ be the ideal of $\esD$ 
	then $\esD$ is integrable if and only if $I_\esD$ is closed under 
	the exterior derivative, {\it i.e.},
	\eqnono{
		dI_\esD \subset I_\esD.
	}
}
\proof{It is sufficient to prove that $d\o \subset I^2_\esD$ for all pfaffian forms in $\O_{{\esD^*}'}$.
	So let $\o\in\O^1_{{\esD^*}'}$ and $X,Y\in\L^1_\esD$ then
	\eqnono{
		d\o(X,Y)=X[\o(Y)]-Y[\o(X)]-\o([X,Y])=-\o([X,Y])
	}
	which is zero for all vector fields in $\L^1_\esD$ if and only if the commutator lies in $\L^1_\esD$, {\it i.e.},
	the distribution is integrable.
}
So we see that we can equivalently express the integrability of the distribution in the co-picture. Now we want to see the structure of the forms
belonging to these subsets of forms, especially those belonging to $\O_{{\esD^*}'}$. So we will make some preliminary definitions.
\defn{sobas}{Let $\esD$ be a $k$-distribution on a riemannian manifold $\ws$, let $\o\in\wedge T^*\ws$ be a differential form
	 on $\ws$ and let $X\in\esD$ be a vector field of the distribution, then we call $\o$ 
	\eqal{3}{
		(i)&\quad{\bf semi-basic},\; if &\qquad&i_X\o=0,\\
		(ii)&\quad{\bf invariant},\; if &\qquad&&\qquad&\Lie_X\o=0,\\
		(iii)&\quad{\bf basic},\; if &\qquad&i_X\o=0,&\qquad&\Lie_X\o=0,
 	}
	$\forall X\in\esD$, with respect to $\esD$. In the case when the form is basic it is also called an absolute integral invariant
	and equivalently satisfies $i_X\o=0,\quad i_X d\o=0$.
}
We now see that the set of semi-basic forms in fact are those forms belonging to $\O_{{\esD^*}'}$. But we also noted that they in general not are closed
under the exterior derivative, not even in the case when $\esD$ is integrable, but we had to introduce the ideal to express the integrability. Now
the set of basic forms do indeed close under the exterior derivative. We can see the difference between semi-basic and basic forms in the case when
the distribution is integrable in the following remark.
\remk{semiandbasic}{Let $\esF$ be a foliation and $\O_{{\esF^*}'}$ be graded algebra of the annihilator of $\esF$, then $\O_{{\esF^*}'}$ is 
	the set of semi-basic forms with respect to $\esF$. Let $\o\in\O^p_{{\esF^*}'}$ be a semi-basic $p$-form, 
	it can then be expressed locally as
	\eqnono{
		\o=\o_{m_1'\cdots m_p'}(x,y)dy^{m_1'}\wedge\ldots\wedge dy^{m_p'}.
	}
	If additionally $\Lie_X\o=0,\;X\in\L^1_{\esF}$, then $\o$ is basic and can then be expressed locally as
	\eqnono{
		\o=\o_{m_1'\cdots m_p'}(y)dy^{m_1'}\wedge\ldots\wedge dy^{m_p'}.
	}
	It should also be pointed out that if $X\in\L_{\esF}$ is a multivector on $\ws$, tangent to the leaves, the basic forms are those forms
	vanishing under the generalized Lie derivative, i.e. $\flie_X\o=0,\quad \forall X\in\L_{\esF}$.
}
So we see that in the integrable case the basic forms are those that are semi-basic and constant along the leaves. As these forms are closed under
the exterior derivative, we can look at cohomology groups on the leaf space.
\defn{basic}{The set of basic forms of a foliation $\esF$ is a subset of $\O_{{\esF^*}'}$ which
	we will denote $\O_{B_\esF}$. The basic forms are closed under the exterior derivative, {\it i.e.},
	\eqnono{
		d: \O_{B_\esF} \longmapsto \O_{B_\esF},\quad \O^p_{B_\esF} \longmapsto \O^{p+1}_{B_\esF}\quad 
	}
	so the basic forms form a subcomplex of the De Rahm complex. We can build the set of closed basic $p$-forms, $Z^p_{B_\esF}$, and
	the set of exact basic $p$-forms, $B^{p}_{B_\esF}$, and form the basic cohomology groups
	\eqal{1}{
		H^p_{B_\esF}:=&Z^p_{B_\esF}/B^{p}_{B_\esF},\\
		H_{B_\esF}:=&\underset{p}{\oplus}H^p_{B_\esF}
	}
	which is the De Rahm cohomology of the leafspace of the foliation.
}
It shall be noted that although the manifold is nice the leaf space need not be. In fact \cite{Co87} argues that in certain cases it is in fact
non-commutative, and the basic cohomology groups can be infinite-dimensional even though $\ws$ is compact. We will not discuss these
basic cohomology groups here but refer to \cite{To88}. We can however say that it is easy to show that $H^1_{B_\esF}\subset H^1(\ws)$.

We will now turn our study to the case when we are given a $p$-form and see what this specific $p$-form can tell us in the sense of distributions.
\defn{kernel}{Let $\o\in\O^p$ be a $p$-form on $\ws$, then the {\bf kernel} of $\o$ and the {\bf rank} of $\o$ at $x\in\ws$,
	denoted $\ker_x\o$ and ${\rm rank}_x\o$ respectively,
	are defined through the kernel and the rank of the map $f_\o|_x:\L_x^1\mapsto\O_x^{p-1}$, defined by
	\eqnono{
		f_\o(X)|_x:=i_{X}\o|_x,\quad X\in\L^1.
	} 
}
Of course a $p$-form does not, in general, be of constant rank, but if it is we simply denote it by rank $\o$. The rank and the kernel is
of course dual to each other in the sense that $ \dim \ker_x\o +{\rm rank}_x\o=m$ where $m$ is the dimension of the manifold, $\ws$. Now
we can, given a specific $p$-form, make the following definition.
\defn{charsub}{Let $\o\in\O$ be a differential form on $\ws$, the {\bf characteristic subspace}, $\esD_x$, of $\o$ at a point
	$x\in\ws$ is defined by
	\eqnono{
		\esD_x:=\ker_x\o \cap \ker_xd\o.
	}
	The {\bf class} of $\o$ at $x$ is the codimension of $\esD_x$ in $T_x\ws$ and
	the {\bf characteristic distribution}, $\esD$, of $\o$ is simply $\esD:=\cup_{x\in\ws}\esD_x$.
}
\remk{class}{The class of a differential form is the smallest number of variables by which we can express it locally.
	If $\o$ is a closed form then the class is equal to its rank.}
To get a little better grip of what the class of a $p$-form is let us consider the four-dimensional Yang--Mills theory.
\exam{first}{Let $F$ be the Lie algebra valued field strength of a abelian gauge potential $A$ in a 4-dimensional space $\ws_4$
	then of course $F$ is a Lie algebra valued two-form on $\ws_4$ and by \cite{Car22} we know that the rank of $F$
	is either $2$ or $4$. If the rank is two we know from definition \ref{charsub} and the fact that $F$ is closed that
	its characteristic distribution, which for an $F$ of constant rank would be a characteristic foliation,
	would be two-dimensional. If this was the case we would for instance know that in a flat manifold
	we could choose coordinates in such a way that the two-dimensional foliation would be global coordinate surfaces.
	Now $F$ would not depend on these coordinates but should effectively be a two-dimensional field strength.
	This is clearly not the case, and this is because the rank of $F$ is in fact four and this is due to the
	self-duality condition $F=*F$.
}
It should be noted that the set of $p$-forms of constant class is of great importance. The reason for this becomes clear by
the following proposition.
\prop{charfoli}{Let $\o\in\O$ be a differential form on $\ws$ with constant class, then the characteristic
	distribution will be of constant dimension and the distribution will be integrable, {\it i.e.}, $\o$
	will define a foliation on $\ws$.}
\proof{Take $X,Y\in\L^1_\esD$ where $\esD$ is the characteristic distribution of $\o$, then
	$\Lie_{[X,Y]}=\Lie_X\Lie_Y\o-\Lie_Y\Lie_X\o=0$, and $i_{[X,Y]}\o=\Lie_X i_Y\o-i_Y\Lie_X\o=0$
	which implies $[X,Y]\in\L^1_\esD$, so $\esD$ is integrable and thus a foliation.}
\cor{sobasic}{Let $\o \in \O$ be a differential form on $\ws$ with constant class, then $\o$ is
	basic with respect to the characteristic foliation defined by $\o$.}
\proof{By definition.}
So a $p$-form of constant class defines a foliation on $\ws$ but then we know that finding a foliation on $\ws$
of dimension $k$ is equal to finding a $p$-form on $\ws$ of constant class $m-k$. Note that the $p$-form need not be a $m-k$ form but
can be an arbitrary $p$-form as long as $p\leq m-k$ and not a $0$-form. 

We look at the $M5$-brane solution.
\exam{M5}{Let $F_4$ be the four-form in $D=11$ supergravity and consider the M5-brane solution to
	the equations of motions in $\ws_{11}$ \cite{Gu92,Du96},
	\eqnono{
		F=\pa^m\Delta(y)\ve_{mnpqr}dy^n \wedge dy^p \wedge dy^q \wedge dy^r
	}
	with coordinates $(x^\m,y^m)$ along the brane and transverse to the brane respectively.
	Then $\ker F=6$ and the rank of $F$ is $5=11-6$. Since $F$ is closed we know that the class of $F$
	is equal to the rank and is therefore equal to $5$. The characteristic distribution of $F$ is nothing
	but the $M5$-brane itself which indeed is integrable and thus defines a foliation of $\ws_{11}$.
	We also see by definition that $F$ is a basic $4$-form and because it is closed it must belong to
	the fourth basic cohomology class, {\it i.e.},
	\eqnono{
		F \in H^4_{B_\esF},
	}
	where $\esF$ is the M5-brane.
	
	For the M2-brane it is instead $*H$ that is basic and closed and thus define the $3$-dimensional foliation of the membrane.}
%%%%%%%%%%%%%%%%%%%%%%%%%%%
%
% Subsection: Distributions on riemannian manifolds
%
%%%%%%%%%%%%%%%%%%%%%%%%%%%
\subsection{Distributions on riemannian manifolds}

We will now proceed to see what structure distributions can have after we have added a metric but first let us introduce notations
regarding mappings with the metric tensor.
\defn{raise}{When considering the metric, $g$, and its inverse as isomorphic mappings from the tangent space into the cotangent space and
	vice versa, $g:T\ws\rightarrow T^*\ws$. We will use the standard musical notation, {\it i.e.},
	\eqnono{
		\li X:=g(X)\in T^*\ws,\quad X\in T\ws,
	}
	\eqnono{
		\ri \vf:=g^{\minus 1}(\vf)\in  T\ws,\quad X\in T^*\ws.
	}
}
We will also need the metric splitting of a two-tensor, so let us introduce notation for this.
\defn{trace}{We will define the metric trace, the anti-symmetrizer and the symmetrizer on (2,0) tensors by
	\eqsnono{
		\tr T&:=&T(\tE_a,\ri\tE^a)\\
		\wedge T(X,Y)&:=&{1\fr 2}(T(X,Y)-T(Y,X))\\
		\odot T(X,Y)&:=&{1\fr 2}(T(X,Y)+T(Y,X))
	}
}
When structuring distributions we need the Levi--Civita connection, {\it i.e.}, the unique metric and torsionfree connection. So let us start by
defining it,  perhaps in an unfamiliar way. 
\defn{lcdel}{Let $\ws$ be a riemannian or pseudo-riemannian manifold with non-degenerate metric, $g$, then the Levi--Civita connection is the unique
	torsionfree connection defined by its action on a 1-form
	\eqnono{
		\del\vf(X,Y):={1\fr 2}(d\vf(X,Y)+\Lie_{\ri\vf}g(X,Y))
	}
}
The more familiar coordinate expression can easily be recovered by taking the coordinate vectors for $X,Y$ and the coordinate differential for $\vf$.

We are now ready to define the deformation tensor related to every distribution on a manifold with metric.
\defn{deform}{Let $\esD$ be a $k$-distribution with projection $\ico$ on a riemannian manifold $\ws$ with non-degenerate metric $g$. Let $\del$ be
	the Levi--Civita connection with respect to this metric and let $\nco:=1-\ico$ be the coprojection of $\esD$. 
	Now define the following tensors with characteristics
	\eqal{2}{
		H,L,K:&\quad&\L^1_\esD\times\L^1_\esD &\longmapsto \L^1_{\esD'}\\
		\k:&&\L^1_{\esD'}&\longmapsto \dsR
	}
	and
	\eqal{3}{
		(i)&&	\quad H(X,Y)	&:=\nco\del_{\ico X}\ico Y	&\quad&\text{\bf deformation tensor},\\
		(ii)&&	L	&:=\wedge H				&&\text{\bf twisting tensor},\\
		(iii)&&	K	&:=\odot H				&&\text{\bf extrinsic curvature tensor},\\
		(iv)&&	\ri\k	&:=\tr H				&&\text{\bf mean curvature tensor},\\
		(v)&&	W	&:=K-{1\fr k}\ri\k \ig			&&\text{\bf conformal curvature tensor}.
	}
	This gives us the decomposition of the deformation tensor in its anti-symmetric, symmetric-traceless and trace parts accordingly,
	\eqnono{
		H=L+W+{1\fr k}\ri\k\ig.
	}
}
These are the definitions of the fundamental tensors of a distribution, and we would like to make some comments about them.
We see that the deformation tensor can be split into the usual anti-symmetric, symmetric-traceless and trace parts. So what does these
parts say? If we just see them as generators of matrix algebras we know that these splitting refers to one tensor, the twisting tensor, that generates
rotations, one tensor that generates deformations but leaves the volume constant, the conformal curvature tensor, and one tensor that scales the
volume, the mean curvature tensor. In the case of distributions this is the same although we now talk about how the distribution
changes while going in normal directions. From some important relations that we will see this becomes evident. But first we need
a relation to prove them.
\lem{liecov}{Let $X,Y,Z\in \L^1$ be vector fields on $\ws$ with metric $g$. Let $\del_X$ be the Levi--Civita connection on $(\ws,g)$ then
	\eqnono{
		(\Lie_X g)(Y,Z)=g(\del_Y X,Z)+g(Y,\del_Z X).
	}
}
\proof{By direct calculation
	\eqal{1}{
		(\Lie_X g)(Y,Z)=&X[g(Y,Z)]-g([X,Y],Z)-g(Y,[X,Z])=\\
				=&X[g(Y,Z)]-g(\del_X Y-\del_Y X,Z)-g(Y,\del_X Z - \del_Z X)=\\
				&\big\{g(\del_X Y,Z)=X[g(Y,Z)]-g(Y,\del_X Z)\big\}\\
				=&g(\del_Y X,Z)+g(Y,\del_Z X).
	}
}
So we get the important relations.
\prop{defrel}{Let $\esD$ be a distribution on a manifold $\ts$ with metric $\tg$, let further $g(X,Y)=\tg(\ico X,\ico Y)$ be the induced metric on the
	distribution, then the symmetric part of the deformation tensor can be written like
	\eqnono{
		K(X,Y)(\vf)=\minus{1\fr 2}\Lie_{\ri\vf'}\ig(X,Y),\;or\quad\li K(X,Y,Z)=\minus{1\fr 2}\Lie_{Z'}\ig(X,Y),
	}
	where the prime denotes projection along the normal directions by $\nco$. The relation for the anti-symmetric part on the other hand is
	\eqnono{
		L(X,Y)={1\fr 2}\nco [\ico X,\ico Y]
	} 
}
\proof{By direct calculation,
	\eqal{1}{
		\minus{1\fr 2}\Lie_{\ri\vf'}\ig(X,Y)=&\minus{1\fr 2}(\tg(\del_{\ico X} \nco \ri \vf,\ico Y)+\tg(\ico X,\del_{\ico Y} \nco \ri \vf))=\\
					=&{1\fr 2}(\tg(\nco \ri \vf,\nco \del_{\ico X} \ico Y)+\tg(\nco \del_{\ico Y} \ico X,\nco \ri \vf))=\\
					=&{1\fr 2}(H(X,Y)(\vf)+H(X,Y)(\vf))\\
					=&K(X,Y)
	}
	and 
	\eqal{1}{
		L(X,Y)=&{1\fr 2}(H(X,Y)-H(Y,X))=\\
			=&{1\fr 2}(\nco \del_{\ico X} \ico Y-\nco \del_{\ico Y} \ico X)=\\
			=&{1\fr 2}\nco [\ico X,\ico Y].
	}
}
Now it is evident that the twisting tensor which can be regarded as rotations of the distributions while going in the normal directions in fact
measures how far the distribution is from being integrable. So we get a natural proposition from this. 
\prop{dist-fol}{Let $\esD$ be a distribution on a manifold, then $\esD$ defines a foliation if and only if $\esD$ is integrable, which on a riemannian manifold
	is equivalent to the vanishing of the tensor $L$ above.}
\proof{$L=0\Rightarrow \nco [\ico X, \ico Y]=0\Rightarrow [X,Y]\in \L^1_\esD, \forall X,Y\in \L^1_\esD \Rightarrow \esD\; integrable.$
	Now \ref{distfoli} completes the proof.}
For the case of the extrinsic curvature we see from \ref{defrel} that it indeed measures the change of the induced
metric on the distribution while going in normal directions. If we look at conformal transformations we can see that the conformal curvature tensor
does not see volume changes.
\prop{defrela}{Let $\ts$ be a riemannian manifold with metric $\tg$, let $I$ be an almost product structure on $\ts$ which split the metric in
	$\tg=\ig+\nlg$ and let $\l=e^{2\phi}$ be a conformal transformation on $\ig$, {\it i.e.}, $\ct\tg=\l\tg$ then the symmetric parts of the 
	deformation tensor will transform like
	\eqsnono{
		\ct K(\vf)&=&K(\vf)+\l^{\minus 1}\ri\vf'[\l]g=K(\vf)+2\ri\vf[\phi]g\\
		\ct\k(X)&=&\k(X)+k\l^{\minus 1} X'[\l]=\k(X)+2kX'[\phi]\\
		\ct W&=&W\\
		\ct L&=&L
	}
}
\proof{By direct calculation.}
If we put all this together we see that we indeed have 8 fundamental classes of distributions on a riemannian manifold.
\defn{umg}{Let $\esD$ be a distribution on a riemannian manifold $\ws$ we have the following 8 different classes
	\eqnono{\begin{tabular}{l|c|c|c|l}
		Name&$L=0$&$W=0$&$\k=0$&Notation\\\hline\hline
		Distribution&&&&$D$\\\hline
		Minimal Distribution&&&x&$MD$\\\hline
		Umbilic Distribution&&x&&$UD$\\\hline
		Geodesic Distribution&&x&x&$GD$\\\hline
		Foliation&x&&&$F$\\\hline
		Minimal Foliation&x&&x&$MF$\\\hline
		Umbilic Foliation&x&x&&$UF$\\\hline
		Geodesic Foliation&x&x&x&$GF$\\\hline
	\end{tabular}}
}
\begin{figure}
\begin{center}
        \includegraphics[width=8cm]{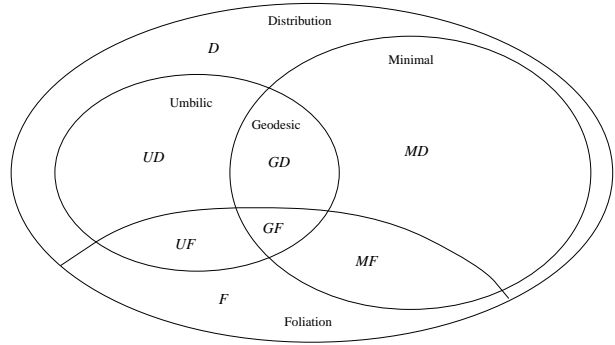} 
\end{center}
\caption{Overview of the different classes of a distribution}
\end{figure}	 
%%%%%%%%%%%%%%%%%%%%%%%%%%%%%%%%%%%%%
%
% Section: Foliations defined by 1-1 tensors
%
%%%%%%%%%%%%%%%%%%%%%%%%%%%%%%%%%%%%%
\section{Foliations defined by (1,1) tensors}

We will in this section show how foliations can be described by certain types of endomorphisms on the tangent bundle. To start with we will therefore review
the concepts of endomorphisms on the tangent bundle. In here we will see that there appears a fundamental tensor known as the Nijenhuis tensor which could
be seen as the curvature of the endomorphism. We will derive this tensor in a different way from the ordinary one. This way of looking at the Nijenhuis tensor
will put it on an equal basis to that of curvatures from connections on fiber bundles.
%%%%%%%%%%%%%%%%%%%%%%%%%%%%%%
%
% Subsection: Basics of endomorphisms
%
%%%%%%%%%%%%%%%%%%%%%%%%%%%%%%
\subsection{Endomorphisms on the tangent bundle}
In this subsection we will look at endomorphisms, which basically are maps from the tangent bundle into itself. These maps can be described by (1,1) tensors
and can equivalently be regarded as maps from the cotangent bundle to itself. We will in this section depend a lot from the results in section 2, {\it i.e.}, we will
need the concepts of generalized Lie derivation and we will need the Nijenhuis-F\"{o}hlicher bracket which plays a central part in the study of the
Nijenhuis tensor. But let us now first define the basic structure of endomorphisms.
\defn{endomorph}{Let $\ws$ be a manifold and $T\ws$ its tangent bundle, then an endomorphism $I$ is a map
	\eqnono{
		\I:T\ws\longmapsto T\ws
	}
	{\it i.e.}, it is a (1,1) tensor acting on the tangent bundle. Let $X\in\L^1$ be vector field on $\ws$ then we
	denote the action of $I$ on $X$ by
	\eqnono{
		I:X\longmapsto I(X)=IX
	}
	which is nothing but the exterior product mapping defined in \ref{exter} {\it i.e.},
	\eqnono{
		\ve_IX=IX
	}
	and can be extended, to an arbitrary $p$-vector, with characteristics
	\eqnono{
		\ve_I:\L\longmapsto\L,\quad\L^p\longmapsto\L^p
	}
	and action for $X_i\in\L^1$
	\eqnono{
		\ve_I(X_1\wedge\ldots\wedge X_p)=\sum_iX_1\wedge\ldots\wedge IX_i\wedge\ldots\wedge X_p
	}
}
We start by noticing that $\ve_I \in {\rm Der}\L$ is a derivation on the set of multivectors on $\ws$. We also know that it maps vectors to vectors,
which immediately lead us to the thought of associating a new bracket to this endomorphism by just taking the commutator of the respective maps.
\defn{ibracket}{The bracket associated with an endomorphism $\I$ is called $I$-bracket and denoted by $[\cdot,\cdot]_I$. It has the characteristics
	of a normal bracket {\it i.e.},
	\eqnono{
		[\;\cdot\;,\;\cdot\;]_I:\L^1\times\L^1\longmapsto\L^1,
	}
	if $X,Y\in\L^1$ are two vector fields it is defined by
	\eqnono{
		[X,Y]_{\I}:=[IX,Y]+[X,IY]-I[X,Y]
	}
	and is thus manifestly antisymmetric. We also see from \ref{genlie} that
	\eqnono{
		[X,Y]_I\equiv\Lie_I(X\wedge Y)
	}
	We see that if $I$ is the identity map the $I$-bracket reduces to the usual bracket. Because of this we will denote
	\eqnono{
		\pa_I:=\Lie_I=[\pa,\ve_I]
	}
	in the case when $I$ is an (1,1) tensor and we see that $\pa_1=\pa$. Now $\pa_I$ can act on a $q$-vector of arbitrary degree,
	in which the characteristics of the map looks like
	\eqnono{
		\pa_I:\L \longmapsto \L, \quad \L^q \longmapsto \L^{q-1}.
	}
}
This new bracket has indeed the properties of a usual vector bracket, {\it i.e.}, it is a anti-symmetric, non-tensorial map taking two vectors into one. The
non-tensoriality looks like
\eqnono{
	[X,fY]_I-f[X,Y]_I=IX[f]Y,
}
and is thus depending on the directional derivative of $f$ along $IX$ instead of $X$ as in the ordinary bracket. Now the original vector bracket is a
Lie bracket, {\it i.e.}, it fulfills the Jacobi identity. One question that immediately arises is if the $I$-bracket is a Lie bracket. Generically the answer to 
this question is no. There are however cases when indeed this $I$-bracket is a Lie bracket, so we need a measure which tells when this is the case,
and this measure will be the Nijenhuis tensor.
\defn{Nijenss}{Let $I$ be an endomorphism on $\ws$, then define the {\bf Nijenhuis tensor} as the failure of the $I$-bracket to be a Lie bracket, {\it i.e.},
	let $X,Y,Z\in\L^1$ be vector fields, then the Nijenhuis tensor is a map with characteristics
	\eqnono{
		\check{N}_I(X,Y,Z):\L^1\times\L^1\times\L^1\longmapsto\L^1,
	}
	so it is a (3,1) tensor and it measures the failure of the $I$-bracket in fulfilling the Jacobi identities. It is defined by
	\eqnono{
		\check{N}_\I(X,Y,Z):=[[X,Y]_\I,Z]_\I+[[Y,Z]_\I,X]_\I+[[Z,X]_\I,Y]_\I
	}
	The Nijenhuis tensor can through the equality
	\eqnono{
		\check{N}_I(X,Y,Z)\equiv\pa_I\circ\pa_I(X\wedge Y\wedge Z)
	}
	be seen as measuring the failure of $\pa_I$ to be a boundary operator.
}
\remk{nijrem}{As $\pa$ is not a well defined boundary operator, $\pa_I$ will of course not be well defined either, not even when it is closed.
	The reason for denoting the Nijenhuis tensor with a check is that it turns up in a more natural way as a (2,1) tensor why we reserve
	the notation $N_I$ to this case.}
It can easily be proved that the Nijenhuis tensor defined above indeed is a tensor, {\it i.e.}, multilinear. We saw that the Nijenhuis tensor measured in what
amount the $I$-bracket failed in fulfilling the Jacobi identities. This is the same as to say that the Nijenhuis tensor measures to what extent the
$I$-bracket fails to be a Lie bracket. The conclusion is that the Nijenhuis tensor measures in what extent the endomorphism $I$ fails to be
a Lie algebra homomorphism on the infinite-dimensional Lie algebra of vector fields on $\ws$. This conclusion will be more transparent when we introduce the
other type of Nijenhuis tensor originating from the treatment of endomorphisms on the cotangent bundle.
\defn{coend}{Let $\ws$ be a manifold and $T^*M$ its cotangent space, let $I$ be an endomorphism of the tangent bundle, then $I^t$ is the natural extension
	characterized by
	\eqnono{
		I^t:T^*\ws\longmapsto T^*\ws.
	}
	Let $\o\in\O^1$, then the action of $I^t$ on $\o$ looks like
	\eqnono{
		I^t:\o\longmapsto I^t(\o)=\o I,
	}
	{\it i.e.}, $I$ acts as a right mapping on $\o$. It is nothing but the interior product of $I$ on a $1$-form and it generalizes as the dual map
	of the exterior product. So for $\o\in\O^p$ we get
	\eqal{1}{
		i_I\o(X_1,\ldots,X_p):&=\o(\ve_I(X_1\wedge,\ldots,\wedge X_p))=\\
					&\sum_i\o(X_1,\ldots,IX_i,\ldots,X_p),
	}
	and we find that $i_I$ is an algebraic derivation of degree $0$ with characteristics
	\eqnono{
		i_I:\O\longmapsto\O,\quad\O^p\longmapsto\O^p
	}
}
We do equally know in this case that $i_I \in {\rm Der}\O$ is a derivation on the cotangent bundle and it will therefore be natural to introduce the
commutator of $i_I$ and the exterior derivative in an analogous way as we introduced the $I$-bracket.
\defn{iextd}{Let $I$ be an endomorphism on a manifold $\ws$ and define the associated exterior derivative, denoted by $d_I$, with characteristics
	\eqnono{
		d_I: \O \longmapsto \O, \quad \O^p \longmapsto \O^{p+1}
	}
	by the commutator
	\eqnono{
		d_{\I}:=[i_I,d]
	}
	which now is the dual map to $\pa_I$. Let $\o\in\O^p$ be a $p$-form, then from \ref{lielie} we see
	\eqsnono{
		d_I\o(X_1,\ldots,X_p)&=&\sum_i(\minus 1)^{i+1}\Lie_{IX_i}\o(X_1,\no{i},X_p)-\\
				&&-\o(\pa_I(X_1\wedge\ldots \wedge X_p))
	}
	or in $I$-bracket notation 
	\eqsnono{
		d_\I\o(X_1,\ldots,X_{p+1})&=&\sum_{i}(-1)^{i+1}\I X_i[\o(X_1,\no{i},X_{p+1})]-\\
				&&-\sum_{i<j}(-1)^{i+j+1}\o([X_i,X_j]_\I,X_1,\no{i}\no{j},X_{p+1})
	}
}
Now we would like to ask whether this new operator, with the same mapping characteristics as the exterior derivative, is a coboundary operator or not, {\it i.e.},
whether it is nilpotent or not. So in analogy to the treatment of the $I$-bracket we introduce a new type of Nijenhuis tensor which measures
to what extent the associated exterior derivative $d_I$ fails in being nilpotent.
\defn{nijen2}{Let $I$ be an endomorphism on $\ws$ and define the {\bf Nijenhuis tensor} as the measure of how much $d_I$ fails to be a coboundary
	operator. The Nijenhuis tensor is thus a (2,1) tensor. Let $X,Y\in \L^1$ be vector fields on $\ws$, then the characteristics of the Nijenhuis
	tensor are
	\eqnono{
		N_I(X,Y): \L^1 \times \L^1 \longmapsto \L^1
	}
	and we define it through the quadratic action of $d_I$ on functions $f \in C^\infty (\ws)$,
	\eqnono{
		<-N_I(X,Y), d f>:=d_\I d_\I f(X,Y).
	}
	As we see the Nijenhuis tensor measures the failure in closure of the operator $d_I$ and can thus be 
	considered as a form of torsion. Alternatively, as the below equivalent definition shows, it measures the
	curvature of the endomorphism, {\it i.e.},
	\eqnono{
		N_\I[X,Y]:=I([X,Y]_\I)-[I(X),I(Y)],
	}
	so the Nijenhuis tensor can be seen as measuring how far this endomorphism is from being a Lie algebra homomorphism
	of the infinite-dimensional Lie algebra of vector fields on $\ws$.
}
\proof{The proof follows from definition \ref{ibracket} and \ref{fronij}.}
\remk{conijen}{Notice that the expression for the Nijenhuis tensor in definition \ref{nijen2} differs by a sign from the original definition. 
	This definition turns out to be more natural in two different aspects. First of all we find that it looks similar to the curvature of 
	algebraic gauge theory and further we see that if $\del$ is a
	connection with torsion on $\ws$, then $\del \wedge \del f(X,Y)=<-T(X,Y),df>$. We will show later that the Nijenhuis tensor can in fact be viewed
	as a kind of torsion, which makes the new sign natural.}
When we write the Nijenhuis tensor on the above form the connection to algebraic gauge theory is clear. In algebraic gauge theory we
have a principal bundle $0 \rightarrow A \rightarrow E \rightarrow B \rightarrow 0$ and a connection $\r: B \rightarrow E$ with curvature
\eqnono{	
	F(X,Y):=\r([X,Y]|_B)-[\r(X),\r(Y)]|_E.
}
The curvature therefore measures to what extent $\r$ fails to be a Lie algebra homomorphism. The conclusion is that the Nijenhuis tensor
describes curvatures in principal bundles. We will look at this more thoroughly later, but first we will examine some basic relations involving
the Nijenhuis tensor that will be needed in the sequel.
We start with a small lemma.
\lem{endo}{Let $I,\; I_1,\; I_2$ be endomorphisms on $\ws$ and let $X,Y\in \L^1$ be two vector fields, then
	\eqal{2}{
		(i)&&\Lie_X\I(Y)&=[X,IY]-I[X,Y]\\
		(ii)&& \quad [I_1,I_2](X,Y)=&[I_1X,I_2Y]+[I_2X,I_1Y]-I_1[X,Y]_{I_2}-I_2[X,Y]_{I_1}
	}
}
\proof{(i) by direct calculation $\Lie_X I (Y)=\Lie_X(IY)-I(\Lie_X Y)=[X,IY]-I[X,Y]$ and\\
	(ii) directly from definition \ref{fronij}.}
\prop{niigen}{Let $I,\; I_1,\; I_2$ be endomorphisms on $\ws$ and let $X,Y\in \L^1$ be two vector fields, then we have the following relations
	involving the Nijenhuis tensor.
	\eqal{2}{
		(i)&&\qquad 	N_I(X,Y)&=(I\Lie_X I-\Lie_{IX}I)(Y)\\ 
		(ii)&&\qquad	N_I&=-{1\fr 2}[I,I]\\
		(iii)&&\qquad	N_{\l\I}&=\l^2N_I\\
		(iv)&&\qquad	N_{I_1+I_2}&=N_{I_1}+N_{I_2}-[I_1,I_2]
	}
}
\proof{(i) follows from the first part of lemma \ref{endo} while (ii), (iii) and (iv) is a direct consequence of the 
	properties of the bracket in lemma \ref{endo}.}
We will also list some properties involving the identity endomorphism which as expected turns out to be trivial.
\prop{nijjjen}{Let $I$ be a endomorphism on $\ws$ and $1$ the identity operator (endomorphism), then we have the following relations involving the 
	Nijenhuis tensor
	\eqal{2}{
		(i)&&\qquad	[1,I]&=0\\
		(ii)&&\qquad	N_{1+I}&=N_{I}
	}
}
\proof{Trivial.}
Now we have defined two types of Nijenhuis tensors, one as the natural one occurring on the space of $p$-vectors and the other
appearing on the space of differential forms. Of course there will be no surprise to us that these two types of tensors in fact
are related. This relation will be seen in following proposition.
\prop{nijnij}{Let $X,\; Y,\; Z\in\L^1$ be vector fields on $\ws$ and let $\tilde{N}_I$ be the Nijenhuis tensor defined in \ref{Nijenss} and 
	$N_I$ be the one defined in \ref{nijen2}. These are then related as
	\eqnono{
		\tilde{N}_\I(X,Y,Z)=\sum_{cycl}N_I([X,Y],Z)+ [N_\I(X,Y),Z]
	}
}
\proof{By definition \ref{fronij} and proposition \ref{genlie} we get
	\eqsnono{
		\tilde{N}_\I(X,Y,Z)&=&\pa_I\pa_I(X \wedge Y \wedge Z)\\
				&=&{1\fr 2}[\vlie_I, \vlie_I](X \wedge Y \wedge Z)\\
				&=&-\vlie_{{1\fr 2}[I,I]}(X \wedge Y \wedge Z)\\
				&=&\vlie_{N_I}(X \wedge Y \wedge Z)\\
				&=&\sum_{cycl}N_I([X,Y],Z)+ [N_\I(X,Y),Z]
	}
}
We also have the complete relation when $d_I$ acts on an arbitrary differential form, which follows.
\prop{disquared}{Let $I$ be an endomorphism on $\ws$ and $d_I$ be its associated coboundary operator, let $\pa_I$ be the formal
	boundary operator associated with the endomorphism, then we have
	\eqsnono{
		d_I d_I \o(X_1, \ldots, X_{p+2}) &=& \sum_{i<j} (\minus 1)^{i+j} \Lie_{N_I(X_i,X_j)} \o(X_1,\no{i}\no{j},X_{p+2})+\\
						&&\o(\pa_I \pa_I (X_1\wedge\ldots\wedge X_{p+2}))
	}
}
\proof{By \ref{fronij} and \ref{lielie}, using $d_Id_I={1\fr 2}[\flie_I,\flie_I]=-\flie_{N_I}$.} 
%%%%%%%%%%%%%%%%%%%%%%%%%%%
%
% Subsubsection: Jordan bracket
%
%%%%%%%%%%%%%%%%%%%%%%%%%%%
\subsubsection{Manifolds with metric}

If we add to the manifold the structure of a non-degenerate metric, we are able to introduce a Levi--Civita connection and 
we can in a similar fashion as above introduce a new bracket structure, the Jordan bracket. 

\defn{jordanbracket}{Let $\ts$ be a riemannian manifold with Levi--Civita connection $\tdel$, then define the {\bf Jordan bracket}, denoted by $\{\cdot,\cdot\}$,
	with the following characteristics:
	\eqnono{
		\{\;\cdot\;,\;\cdot\;\}: \L^1\times\L^1 \longmapsto \L^1,
	}
	by
	\eqnono{
		\{X,Y\}:=\tdel_XY+\tdel_YX
	}
	where $X,Y\in\L^1$ are vector fields on $\ts$. Now define the Jordan bracket associated with an endomorphism $I$, denoted $\{\cdot,\cdot\}_I$,
	in an analogous fashion to $[\cdot,\cdot]_I$, by
	\eqnono{
		\{X,Y\}_I:=\{IX,Y\}+\{X,IY\}-I\{X,Y\}.
	}
}
We see that the Jordan bracket associated with an endomorphism is defined in a similar fashion as the $I$-bracket was earlier. It should also be
pointed out that the Jordan bracket and the usual vector bracket of two vectors, $X,Y\in\L^1$ in fact only measure the symmetric 
and anti-symmetric parts of the tensor $\del_XY$. We can also introduce the Jordan tensor in the same way as we did with the Nijenhuis tensor.
\defn{jordantensor}{Let the triplet $(\ts,\tg,I)$ define a riemannian almost product structure, and let $\{\cdot,\cdot\}$ be the Jordan bracket,
	then define the {\bf Jordan tensor} associated to $I$, denoted $M_I$, with the following characteristics:
	\eqnono{
		M_I: \L^1\times\L^1 \longmapsto \L^1,
	}
	by
	\eqnono{
		M_I(X,Y):=I\{X,Y\}_I-\{IX,IY\}
	}
	where $X,Y\in\L^1$ are vector fields on $\ts$. The analogy to the Nijenhuis tensor is obvious .
}
So we see that the Jordan tensor measures the failure of the Jordan bracket to commute with the endomorphism, $I$. We also get similar relations 
for the Jordan tensor as for the Nijenhuis tensor earlier.
\prop{jordantensorrelations}{Let $I,\; I_1,\; I_2$ be endomorphisms on $\ts$ and let $X,Y\in \L^1$ be two vector fields. Introduce
	the operator $\scT_X:=\tdel_X-\Lie_X$, then we have the following relations
	involving the Jordan tensor.
	\eqal{2}{
		(i)&&\qquad 	M_I(X,Y)&=(I\scT_X I-\scT_{IX}I)(Y)\\ 
		(ii)&&\qquad	M_I&=-{1\fr 2}\{I,I\}\\
		(iii)&&\qquad	M_{\l\I}&=\l^2M_I\\
		(iv)&&\qquad	M_{I_1+I_2}&=M_{I_1}+M_{I_2}-\{I_1,I_2\}
	}
}
In short, the Nijenhuis tensor measures the non-commutativity between an endomorphism $I$ and the antisymmetric part of the 
Levi--Civita connection, while the Jordan tensor measures the non-commutativity between an endomorphism $I$ and the symmetric part of the Levi--Civita
connection. As the anti-symmetric part of the Levi--Civita connection is nothing but the usual vector bracket (or the exterior derivative if seen as
acting on forms), we note that the Nijenhuis tensor is independent of the metric and thus definable even without a metric present. This has been commented on earlier.
If we now have a metric, the two structures can be combined naturally to form the deformation tensor associated with an endomorphism.
\defn{Ideformation}{Let $I$ be an endomorphism on a manifold $\ws$ with non-degenerate metric, $g$. Let $\del$ be the Levi--Civita connection on $\ws$ and 
	define the {\bf deformation tensor} associated with
	the endomorphism $I$, denoted $H_I$, with the following characteristics:
	\eqnono{
		H_I:\L^1 \times \L^1 \longmapsto \L^1.
	}
	$H_I$ is defined by the expression
	\eqnono{
		H_I(X,Y):=(I\del_X I-\del_{IX}I)(Y),
	}
	where $X,Y\in \L^1$ are two vector fields on $\ws$. We immediate note the equivalent definition
	\eqnono{
		H_I(X,Y):=N_I(X,Y)+M_I(X,Y).
	}
}
We will later see that in the case where the endomorphism $I$ is a riemannian almost product structure the deformation tensor will be analogous
to that in the earlier section.
%%%%%%%%%%%%%%%%%%%%%%%%%%%
%
% Subsection: Foliations from endomorphisms
%
%%%%%%%%%%%%%%%%%%%%%%%%%%%
\subsection{Foliations from endomorphisms}

The preceding sections give us the opportunity to formulate the concepts of distributions and foliations in the framework of
a special type of endomorphism on the tangent bundle. We will see that the type of endomorphism will be very similar to that of an
almost complex structure. But to start with we will change our notation a bit in order to get a more compact language when considering
distributions on a manifold.
\nota{nynot}{We will denote the objects on our space with an underline, {\it i.e.},
	\eqal{2}{
		&\ts&	&\text{Manifold}\\
		&T\ts&	&\text{Tangent bundle of $\ts$}\\
		&T^*\ts\qquad&&\text{Cotangent bundle of $\ts$}\\
		&\tg&	&\text{Metric on $\ts$}\\
		&\td&	&\text{Exterior derivative}\\
		&\ul{X}&&\text{Vector field on $\ts$}
	}
	to list the primarily used objects. We will use this underlining principle for all objects on $\ts$ whenever there may be a risk of confusion.
}
When considering endomorphisms in the preceding subsection, where we defined the Nijenhuis tensor, we were treating endomorphisms in the most general sense
and had no conditions on the endomorphism $I$ at all. But there are of course certain types of endomorphisms that are more interesting than
others. In mathematics there are four basic types which are of great importance. We will define them below.  
\defn{11ten}{Let I be an endomorphism tensor of type (1,1), {\it i.e.}, it maps $T\ts\rightarrow T\ts$ or $T^*\ts\rightarrow T^*\ts$ then
	I is called
	\eqal{3}{
		(i)&\qquad&	&{\bf Nilpotent}, if \qquad&&I^2=0,\\
		(ii)&\qquad& 	&{\bf Idempotent}, if &&I^2=I,\\
		(iii)&\qquad&	&\text{{\bf Almost product structure}}, if \qquad&&I^2=1,\\
		(iv)&\qquad&	&\text{{\bf Almost complex structure}}, if \qquad&&I^2=\minus 1.
	}
}
Of course the concepts nilpotent and idempotent could be generalized to hold for a different power than 2, but otherwise these are the four basic types.
Interesting to note is that for a nilpotent endomorphism $\ker I \subset {\rm Im} I$ which implies that ${\rm rank}I\leq [n/2]$. For an idempotent endomorphism
the rank can be arbitrary. The last two types of endomorphisms which are called almost product (complex) structures are both of full rank.
In this section we will see that an almost product structure will be just the kind of endomorphism that one needs in the theory of distributions and 
foliations. Although the study of almost product structures could take place without introducing a metric on the manifold, we will focus
on the treatment of manifolds with a metric. We will only point out that, as seen in previous subsection, all structure involving only the Nijenhuis tensor
exist even without metrics. But let us now introduce a metric on the manifold.
\defn{i2}{Let I be an almost product structure on a manifold $\ts$ with riemannian metric $\tg$ and let $X,\;Y\in T\ts$ be vector fields. Then the
	triplet ($\ts$, $\tg$, $I$) is called an riemannian almost product structure if
	\eqnono{
		\tg(IX,IY)=\tg(X,Y)
	}
	or in other words, $I$ is a automorphism of $\tg$ in the sense that the following diagram commutes:
	\eqnono{
		\xy
			\xymatrix{T\ts\ar[d]_{\tg}\ar[r]^I&T\ts\ar[d]^{\tg}\\
				T^*\ts\ar[r]_{I^t}&T^*\ts}
		\endxy
	}
	{\it i.e.},
	\eqnono{
		I^t\circ\tg\circ\I=\tg
	}
}
So we see that the endomorphism $I$ conserves the length of a vector. This immediately tells us that it must be a local $O(\tm)$ transformation
on the tangent bundle. In the above definition we required that the metric satisfied $\tg(X,Y)=\tg(IX,IY)$ but of course from any riemannian metric
not satisfying this we can always construct a new one just by taking $\tilde{\tg}(X,Y):=\tg(X,Y)+\tg(IX,IY)$ which would then satisfy the above condition.
Therefore, this implies no restriction on the manifold.

We will now look at the properties of a riemannian almost product structure in a little more detail to find out in what sense it defines distributions
on the manifold.
\prop{i1}{Let the triplet $(\ts,\tg,I)$ define a riemannian almost product structure on $\ts$ with $\dim \ts=\tm$, then 
	\rnum{
		\item I is a local $O(\tm)$ matrix with ${1\fr 2}\tm(\tm-1)$ independent components.
		\item All eigenvalues are $\pm1$.
		\item $\tr I=2k-\tm$, where $k$ is the number of positive eigenvalues.
		\item There is a preferred base called the oriented base in which $I$ is diagonal and ordered, {\it i.e.}, it takes the form
			\eqnono{
				I=\mx{cccccc}{1&&&&&\\
						&\ddots&&&&\\
						&&1&&&\\
						&&&\minus 1&&\\
						&&&&\ddots&\\
						&&&&&\minus 1\\}
			}
	}
}
\proof{(i) Let $\{E_\ta\}$ be an orthonormal frame, then $\eta_{\ta\tb}=\tg(E_\ta,E_\tb)=\tg(IE_\ta,IE_\tb)=I_\ta{}^\tc I_\tb{}^\td\eta_{\tc\td}
	\Rightarrow I\in O(\tm)$.
	(ii) $1=I^2=PDP^{\minus 1}PDP^{\minus 1} \Rightarrow D^2=1 \Rightarrow$ all eigenvalues $\pm 1$.
	(iii) $\tr I=k+(\minus 1)(\tm-k)=2k-\tm$.
	(iv) Let $\{E_\ba=(E_a,E_{a'})\}$ be a oriented and ordered base, then $I=E^a E_a - E^{a'}E_{a'}$.
}
These properties of an almost product structure tells us that if we express our vectors in terms of the eigenvectors of $I$ (preferably in
the oriented base) $I$ acts as reflecting the vectors in the hyperplane spanned by the eigenvectors with positive eigenvalue 1. Now the set of vectors
lying in this hyperplane will be invariant under $I$ while those lying in the normal hyperplane will change to the opposite direction under $I$. One
can say that $I$ breaks the structure group $O(\tm)$ of $T\ts$ down to $O(k)\times O(k')$ where $k':=\tm-k$. In that sense the set of almost product structures
with $k$ positive eigenvalues is parameterized by the grassmannian,
\eq{
	I \in Gr(k,\tm)={O(\tm) \fr {O(k) \times O(k')}}
}
The grassmannian has $kk'=k(\tm-k)$ independent components and parameterizes the space of $k$-planes in $\dsR^\tm$. 
We can now let an almost product structure define two complementary distributions for us by taking these complementary hyperplanes
spanned by the eigenfunctions with positive eigenvalues and by the eigenfunctions with negative eigenvalues respectively. 
\defn{ideffol}{Let I be an almost product structure on $\ts$, then $I$ defines two natural distributions of $T\ts$, denoted $\esD$ and $\esD'$ respectively, 
	in the following way. Let
	\eqsnono{
		\esD_x&:=&\{X\in T_x\ts:IX=X\},\\
		\esD'_x&:=&\{X\in T_x\ts:IX=\minus X\},\\
	}
	then
	\eqnono{
		\esD:=\bigcup_{x\in\ts}\esD_x,\quad\esD':=\bigcup_{x\in\ts}\esD'_x.
	}
}
Again it should be noted that the distributions defined above are independent of the existence of a metric on $\ts$. The main difference is that
in the case where we have a metric the structure group of $T\ts$ breaks down from $GL(\tm)$ down to $O(\tm)$, and the almost product structure
will thus be parameterized under the grassmannian space previously introduced. In the case where we don't have a metric the almost product structure
would be parameterized under $GL(\tm)/(GL(k)\times GL(k'))$. As we know that the almost product structure squares to one, we can define two complementary
projection operators, which of course also are endomorphisms on the tangent bundle.
\defn{projs}{From an almost product structure $I$ on a manifold $\ts$ we can define two projection operators through
	\eqsnono{
		\ico&:=&{1\fr 2}(1+I)\\
		\nco&:=&{1\fr 2}(1-I).\\
	}
	These will be mappings in the sense $\ico:T\ts\rightarrow \esD$ and $\nco:T\ts\rightarrow \esD'$ respectively.
}
We see that we can regard the distributions $\esD,\esD'$ as subbundles of the tangent bundle. In this sense the projection operators take an element
in $T\ts$ down to an element in $\esD$ and $\esD'$ respectively. The map is by definition surjective and if we require that the almost product structure
is $C^k$, or even $C^\infty$, the map will be a surjective submersion. We can also introduce canonical inclusions with respect to these submersions.
\defn{incl}{We can define the canonical inclusions $\tilde{\ico}$ and $\tilde{\nco}$ of $\esF$ and $\esF'$ in $T\ts$ by the commutativity of the following
	diagram:
	\eqnono{
		\xy
			\xymatrix{\esD\ar[r]^{\tilde{\ico}}\ar[d]_{Id}&T\ts\ar[dl]^{\ico}\ar[dr]_{\nco}&\esD'\ar[l]_{\tilde{\nco}}\ar[d]^{Id}\\
				\esD&&\esD'}
		\endxy
	}
}
We see from the definition that these inclusions are equivalently defined by $\ico \tilde{\ico}=1_\esD,\nco \tilde{\nco}=1_{\esD'}$. 
These projection operators split the tangent bundle into two complementary parts. 
\prop{exa}{Let $\ts$ be a manifold, $T\ts$ its tangent bundle, $\esF\subset T\ts$ and $\esF'\subset T\ts$ two subbundles with projectors $\ico$ and $\nco$ 
	respectively, then the following statements are equivalent:
	\rnum{
	\item	$\esD\cap \esD'=0,\quad \esD\cup \esD'=T\ts$
	\item	The short sequence
		\eqnono{\begin{CD}
			0 @>>> \esD @>\tilde{\ico}>> T\ts@>\nco >>\esD'@>>> 0
		\end{CD}} is exact.
	}
}
\proof{Trivial by the exactness of the sequence $\esF'=T\ts/\esF$.}
What this tells us is in fact that the almost product structure $I$ in form of its associated projection operators splits the tangent bundle into
\eq{
	T\ts=\esD\oplus \esD'.
}
Later we will show that in the case of a principal bundle,
one of the projection operators of $I$ will in fact be the connection of the principal bundle, and the Nijenhuis tensor, $N_I$, will 
measure the curvature of this connection. An interesting point regarding these inclusion maps is that if we restrict the projection operators
to a submanifold $\ws$ of $\ts$ in such way that $T\ws$ is spanned by the eigenvectors of $\ico$, and that these furthermore are integrable, then a map
$f:\ws \mapsto \ts$ is in fact an embedding and $\tilde{\ico}$ the associated embedding matrix. We will see this more transparently later, but
first we will come back to the Nijenhuis tensor and investigate how its structure is affected by imposing the condition of an almost product structure to 
the endomorphism $I$.
\lem{nijenigen}{Let $I$ be an almost product structure on a manifold $\ts$ and let its associated projection operators be
	$\ico:={1\fr 2}(1+I)$, $\nco:={1\fr 2}(1-I)$, then
	\eqal{2}{
		(i)&&\qquad	N_{\ico}=&N_{\nco}\cr
		(ii)&&		N_I=&4N_{\ico}\cr
		(iii)&&		{1\fr 2}[\ico,\nco ]=&N_{\ico}\cr
		(iv)&&\qquad	N_{\ico}(X,Y)=&-\nco [\ico X,\ico Y]-\ico [\nco X, \nco Y]
	}
}
\proof{By direct calculation\\ 
	(i) $N_{\ico}={1\fr 2}[\ico, \ico]={1\fr 2}[1-\nco, 1-\nco]={1\fr 2}[\nco, \nco]=N_{\nco}$,\\
	(ii) $N_I=N_{\ico-\nco}=N_{2\ico-1}=4N_{\ico}$,\\
	(iii) ${1\fr 2}[\ico,\nco]={1\fr 2}[\ico,1-\ico]=N_{\ico}$,\\
	(iv) -\eqal{1}{N_{\ico}(X,Y)=&[\ico X,\ico Y]+\ico^2 [X,Y]-\ico [\ico X,Y]-\ico [X,\ico Y]=\\
			=&[\ico X,\ico Y]+\ico [X,Y]-\ico [X,Y]+\ico [\nco X,Y] -\ico [X,\ico Y]=\\
			=&\nco [\ico X,\ico Y]+\ico [\ico X,\ico Y]+\ico [\nco X,\ico Y]+\\
			&+\ico [\nco X,\nco Y]-\ico [\ico X,\ico Y]-\ico [\nco X,\ico Y]=\\
			=&\nco [\ico X,\ico Y]+\ico [\nco X,\nco Y]}}
From the above lemma it is clear that the Nijenhuis tensor measures to what extent the two complementary distributions, associated with an
almost product structure, fail to integrable.
\prop{nijol}{Let the triplet $(\ts,\tg,I)$ define an riemannian almost product structure and let $L,\; L'$ be the twisting tensors of the distributions defined by $I$,
	then
	\eqnono{
		{1\fr 8}N_I=-L-L'.
	}
}
\proof{By \ref{defrel} and \ref{nijol}.}
Again it is noted that all these tensors are invariant under the metric and exist even without a metric on the manifold. We see that in case we have
a foliation, proposition \ref{dist-fol} tells us that at least one of the twisting tensors vanishes. This is the same as saying that $N_I(X,Y)$ is an
eigenvector of $I$, {\it i.e.}, $IN_I(X,Y)=\pm N_I(X,Y)$. In the case when both associated distributions are integrable, the Nijenhuis tensor vanishes and
we have two complementary foliations on the manifold. We will see later that this will lead us to the case where the exterior algebra in fact splits
and becomes doubly graded under the exterior derivatives associated with complementary projections $\ico,\nco$ of $I$. But let us first see
what extra structure an almost product structure will give us in the case where we indeed have a metric on the manifold. We start be noticing that
to every metric $\tg$ on $\ts$, we have two complementary metrics associated with the almost product structure.
\defn{gandg}{Let $\ts$ be a riemannian or pseudo-riemannian manifold with metric, $\tg$, $I$ a reflective structure with $\ico$ and $\nco$ the
	corresponding projectors, then define the two associated metrics with respect to the reflective structure by
	\eqnono{
		\ig(X,Y):=\tg(\ico X,\ico Y),\quad \nlg(X,Y):=\tg(\nco X,\nco Y)
	}
	which implies that $\tg$ splits into these two parts, {\it i.e.},
	\eqnono{
		\tg=\ig+\nlg.
	}
}
Of course we note that it is the condition $\tg(IX,IY)=\tg(X,Y)$ that implies that two complementary vectors are orthogonal, {\it i.e.}, $\tg(\ico X,\nco Y)=0$.
In an analogous way to which we deduced the new structure to the Nijenhuis tensor we can proceed to find out how an almost product structure
reduces the Jordan tensor. We will see that due to the similar bracket structure the structure found in the Nijenhuis tensor will be similar.
\lem{almostjordan}{Let $I$ be an almost product structure on a manifold $\ts$ and let its associated projection operators be
	$\ico:={1\fr 2}(1+I)$, $\nco:={1\fr 2}(1-I)$. Let $M$ denote the Jordan tensor, then
	\eqal{2}{
		(i)&&\qquad	M_{\ico}=&M_{\nco}\cr
		(ii)&&		M_I=&4M_{\ico}\cr
		(iii)&&		{1\fr 2}\{\ico,\nco \}=&M_{\ico}\cr
		(iv)&&\qquad	M_{\ico}(X,Y)=&-\nco \{\ico X,\ico Y\}-\ico \{\nco X, \nco Y\}
	}
}
\proof{Similar to that of \ref{nijol}.}
As the Jordan bracket is just the symmetric part of the covariant derivative while the usual vector bracket can be regarded as the anti-symmetric part,
the structure on the bracket level will of course be similar, but they do of course measure two different things. Notable is that, in contrary to the Nijenhuis
tensor, the Jordan tensor makes no sense in a manifold without metric but the connection used in the Jordan bracket must be metric. 
\prop{almostjordan2}{Let the triplet $(\ts,\tg,I)$ define a riemannian almost product structure, $K,\; K'$ be the extrinsic curvature tensors 
of the distributions defined by $I$,
	then
	\eqnono{
		{1\fr 8}M_I=-K-K'.
	}
}
\proof{By lemma \ref{almostjordan} and definition \ref{deform}.}
So, put together, we see that all the structure of two complementary distributions can be put into this single almost product structure $I$. The
deformation tensors are recovered by the associated Nijenhuis tensor and Jordan tensor, of which the Nijenhuis tensor contains the integrability
conditions while the Jordan tensor contains the extrinsic curvature parts. We will use this in the classification scheme of riemannian almost product
structures in next subsection. Now however we are interested in the connection $\tdel$. First of all it is easily proven that although annihilating
the metric $\tg$ it does not annihilate $\ig,\nlg$. We would then instead like a new connection which we will denote $\adel$ and call the adapted
connection that does annihilate all these metrics so it commutes with all of them.
\defn{adap}{Let $\ts$ be a riemannian or pseudo-riemannian manifold with non-degenerate metric $\tg$ and corresponding Levi--Civita connection $\tdel$.
	Let $\I$ define a foliation in the previous sense, then the following two definitions of the adapted connection are equivalent
	\rnum{
	\item	$\adel_XY:=\tdel_XY+A(X,Y), \quad A(X,Y):={1\fr 2}\I\tdel_X\I(Y)$
	\item	$\adel_XY:=\ico\tdel_X\ico Y+\nco\tdel_X\nco Y$
	}
}
\proof{The proof is immediate.}
We will soon prove that this connection indeed annihilates all the metrics, and it is therefore suitable for calculations on the subbundles generated
by the almost product structure $I$. But first we will introduce yet another connection, called the Vidal connection \cite{Vi73}, not with the property
of being metric but with additional properties which will become clear later.
\defn{rdel}{Let $\ts$ be a riemannian or pseudo-riemannian manifold with non-degenerate metric $\tg$ and corresponding Levi--Civita connection $\tdel$,
	let $\I$ define a foliation in the previous sense, then the Vidal connection is defined by
	\eqnono{
		\rdel_XY:=\adel_XY+B(X,Y),\quad B(X,Y):={1\fr 4}(\tdel_{IY}I+I\tdel_YI)(X).
	}
}
Of course both the tensors $A$ and $B$ will only contain parts of the deformation tensor and are in fact related.
\prop{boa}{Let $B$ be the tensor defined in \ref{rdel} and $A$ the tensor defined in \ref{adap}, then we can express the tensor $B$ in terms of $A$
	and the almost product structure $I$ as
	\eqnono{
		B(X,Y)={1\fr 2}\left(A(Y,X)-A(IY,IX)\right).
	}
}
\proof{By direct calculation from the definitions,
	\eqal{1}{
		{1\fr 2}\left(A(Y,X)-A(IY,IX)\right)=&{1\fr 4}\left(I\tdel_YI(X)-I\tdel_{IY}I(IX)\right)=\\
							=&{1\fr 4}(I\tdel_YI+\tdel_{IY}I)(X)=\\
							=&B(X,Y)
	}
}
We will see the total structure of these two connections last in this subsection, and we will first list a number of their fundamental properties.
The most important property that both these connections satisfy is that they commute with the almost product structure.
\prop{aroni}{Let $\adel$ denote the adapted connection defined in \ref{adap} and $\rdel$ the Vidal connection defined in \ref{rdel}, then
	their principal feature is that they both commute with the almost product structure $I$, {\it i.e.},
	\eqnono{
		\adel_XI=\rdel_XI=0
	}
}
\proof{(i) \eqal{1}{
		\adel_X IY=&\tdel_X IY+{1\fr 2}(I\tdel_X I)(IY)=\\
			=&I\tdel_X Y+(\tdel_XI)(Y)-{1\fr 2}(\tdel_XI)(Y)=\\
			=&I\tdel_XY+{1\fr 2}I^2(\tdel_X I)(Y)=I\adel_XY}
	(ii) \eqal{1}{
		\rdel_X IY=&\adel_X IY+{1\fr 4}(\tdel_{I^2Y}I+I\tdel_{IY}I)(X)=\\
			=&I\adel_XY+{1\fr 4}(I^2\tdel_{Y}I+I\tdel_{IY}I)(X)=\\
			=&I\rdel_XY}
}
As already mentioned, the adapted connection is metric. This is not the case for the Vidal connection, but it is nevertheless important.
Its properties will be examined in the following subsection, where the basic types of riemannian almost product structures will be classified. But let us
now show that the adapted connection indeed annihilates all associated metrics. To help us out we need the following lemma.
\lem{symadel}{Let $A$ be the tensor defined in \ref{adap}, then we have the relation
	\eqnono{
		\tg(A(X,Y),Z)+\tg(Y,A(X,Z))=0
	}
}
\proof{$2\tg(A(X,Y),Z)=\tg(I\tdel_XI(Y),Z)=\tg(\tdel_XI(Y),IZ)=$\\
	$\tg(\tdel_XIY-I\tdel_XY,IZ)=-\tg(Y,I\tdel_XIZ-\tdel_XZ)=-\tg(Y,I\tdel_XI(Z))=$\\
	$-2\tg(Y,A(X,Z))$}
Now it is straightforward to prove that the adapted connection is metric.
\prop{adelmetric}{Let the triplet $(\ts,\tg,I)$ be a riemannian almost product structure on $\ts$ and $\adel$ the adapted connection
	defined in \ref{adap}, then this connection is metric with respect to the splitting of $\tg$ according to \ref{gandg}, {\it i.e.},
	\eqal{1}{
		\adel \ig&=0\\
		\adel \nlg&=0
	}
}
\proof{We have 
	\eqal{1}{
		\adel_X\tg(Y,Z)=&X[\tg(Y,Z)]-\tg(\adel_XY,Z)-\tg(Y,\adel_XZ)=\\
			=&-\tg(A(X,Y),Z)-\tg(Y,A(X,Z))=0,
	}
	and $\adel_XI=0 \Rightarrow \adel_X \ico=0$ so we see
	\eqal{1}{
		(\adel_X\ig)(Y,Z)=&X[\ig(Y,Z)]-\ig(\adel_XY,Z)-\ig(Y,\adel_XZ)=\\
			=&X[\tg(\ico Y,\ico Z)]-\tg(\ico \adel_XY,\ico Z)-\tg(\ico Y,\ico \adel_XZ)=\\
			=&X[\tg(\ico Y,\ico Z)]-\tg( \adel_X\ico Y,\ico Z)-\tg(\ico Y, \adel_X\ico Z)=\\
			=&(\adel_X\tg)(\ico Y,\ico Z)=0}
}
We can now see in a more transparent way how the different parts of these connections look. It becomes most conceptually clear if we use the oriented base. 
\prop{del}{Let the triplet $(\ts,\tg,I)$ define a riemannian almost product structure, let $\ul{\o}$, $\tilde{\ul{\o}}$ and $\Tilde{\Tilde{\ul{\o}}}$
	denote the connection one-forms of the Levi--Civita connection, the adapted connection and the Vidal connection respectively.
	Let furthermore $H,H'$ denote the deformation tensors with respect to $I$ and $C,C'$ be coefficients of anholonomy, then
	\eqsnono{
		\ul{\o}&=&\left[\mx{cc}{\o&H\\-H^t&\O},\mx{cc}{\O'&H'\\-{H'}^t&\o'}\right]\\
		\tilde{\ul{\o}}&=&\left[\mx{cc}{\o&\;\;0\;\;\\\;\;0\;\;&\O},\mx{cc}{\O'&\;\;0\;\;\\\;\;0\;\;&\o'}\right]\\
		\tilde{\tilde{\ul{\o}}}&=&\left[\mx{cc}{\o&\;\;0\;\;\\\;\;0\;\;&C},\mx{cc}{C'&\;\;0\;\;\\\;\;0\;\;&\o'}\right]\\
	}
}
\proof{Let $E_{\ba}=(E_a,E_{a'})$ be the normed eigenvectors of $I$, {\it i.e.}, $IE_a=E_a,IE_{a'}=-E_{a'}$, then we get $\ul{\o}$
	from the definition $\del_{\ba}E_{\bb}=:\ul{\o}_{\ba\bb}{}^{\bc}E_{\bc}$ and the definition of the deformation tensor 
	$H_{ab}{}^{c'}:=\nco \tdel_aE_b=\o_{ab}{}^{c'}E_{c'}$. We have furthermore denoted the normal connections by $\O$, {\it i.e.}, 
	$\ul{\o}_{ab'}{}^{c'}=:\O_{ab'}{}^{c'}$. Now from the relation $A(X,Y)={1\fr 2}(I\tdel_XIY-\tdel_XY)$ we get
	in the same basis
	\eqnono{
		A_{\ba\bb}{}^\bc=\left[\mx{cc}{0&-H_{ab}{}^{c'}\\-H_{ab'}{}^c&0},\mx{cc}{0&-{H'}_{a'b'}{}^c\\-{H'}_{a'b}{}^{c'}&0}\right]
	}
	so $\tilde{\ul{\o}}$ follows. If we write $B(X,Y)={1\fr 4}(\tdel_{IY}Ix-I\tdel_{IY}X+I\tdel_YIX-\tdel_YX)$ we similarly get
	\eqnono{
		B_{\ba\bb}{}^\bc=\left[\mx{cc}{0&0\\0&-{H'}_{b'a}{}^{c'}},\mx{cc}{-H_{ba'}{}^c&0\\0&0}\right].
	}
	Finally, from the torsion equation we have 
	$0=\o_{ab'}{}^{c'}-\o_{b'a}{}^{c'}-C_{ab'}{}^{c'}\Rightarrow \O_{ab'}{}^{c'}-H'_{b'a}{}^{c'}=C_{ab'}{}^{c'}$.
}
%%%%%%%%%%%%%%%%%%%%%%%%%%
%
% Subsection: Classification of almost product manifolds
%
%%%%%%%%%%%%%%%%%%%%%%%%%%
\subsection{Almost product manifolds, the classification}

In this section we will present the different classes of riemannian almost product structures, which will be shown to follow from the different
classes of deformation tensors of section (3). These different classes are primarily split into three different types,
basically refering to the three cases when either both distributions associated with an almost product structure are integrable, only one is, or the
last type where none is integrable. In the first case the manifold is doubly foliated, in the second singly, and in the third not foliated at all,
of course with respect to the almost product structure.

%%%%%%%%%%%%
%
% Subsubsection: Three types from Nijenhuis tensor.
%
%%%%%%%%%%%%
\subsubsection{The types defined by the Nijenhuis tensor}

To begin with we will see that there are relations between the Nijenhuis tensor and the two new connections introduced in the preceding
subsection. These relations are characterized by only involving the torsion parts of the two connections.
\prop{ator}{Let the triplet $(\ts,\tg,I)$ define an riemannian almost product structure, let $N_I$ denote the Nijenhuis tensor of $I$ and 
	$\adel$ the adapted connection defined in \ref{adap}, then we have the following relation
	\eqnono{
		{1\fr 2}N_I(X,Y)=\aT(X,Y)+\aT(IX,IY)
	}
}
\proof{By definition \ref{adap} we get
	\eqal{1}{
	\!\!\!\!\!\!\!\!\aT(X,Y)+&\aT(IX,IY)=\adel_XY - \adel_YX - [X,Y] + \adel_{IX}IY - \adel_{IY}IX - [IX,IY]=\\
				=&{1\fr 2}(\tdel_XY+I\tdel_XIY - \tdel_YX-I\tdel_YIX)-[X,Y]+\\
				&+{1\fr 2}(\tdel_{IX}IY + I\tdel_{IX}Y - \tdel_{IY}IX-I\tdel_{IY}X)-[IX,IY]=\\
				=&{1\fr 2}(I\tdel_XIY - I\tdel_YIX + I\tdel_{IX}Y - I\tdel_{IY}X-I^2[X,Y]-[IX,IY])=\\
				=&{1\fr 2}(I[X,IY]+I[IX,Y]-I^2[X,Y]-[IX,IY])=\\
				=&{1\fr 2}N_I(X,Y).
	}
}
In the case of the Vidal connection, which in the classification scheme will be more important to us, we have an even stronger relation.
\prop{rtor}{Let the triplet $(\ts,\tg,I)$ define an riemannian almost product structure, let $N_I$ denote the Nijenhuis tensor of $I$ and 
	$\rdel$ denote the Vidal connection defined in \ref{rdel}, then we have the relation
	\eqnono{
		{1\fr 4}N_I(X,Y)=\rT(X,Y)
	}
}
\proof{
	Similar to the proof in \ref{ator}.
}
We see that the torsion of the Vidal connection has a one-to-one correspondence with the Nijenhuis tensor while in the case of the adapted connection
the torsion tensor contains the Nijenhuis tensor but also some additional terms. These terms are in fact the entire deformation tensor, so
the adapted connection is not very suitable for our study of different riemannian almost product structures. Now putting all information involving
the Nijenhuis tensor together, we get the following theorem.
\thm{intcoint}{Let the triplet $(\ts,\tg,I)$ define a riemannian almost product structure, let $\esD,\esD'$ be the associated distributions and let
	$L,L'$ be the skew tensors of the deformation, then the first type of almost product structure corresponding to 
	a doubly foliated manifold can be seen by the following equivalent statements,
	\eqal{2}{
		(i)&&\qquad	&N_I=0,\\
		(ii)&&\qquad	&L=0,\quad L'=0,\\
		(iii)&&\qquad	&\rdel\quad \text{is torsionless},\\
		(iv)&&\qquad	&\esD,\esD'\quad \text{are integrable}.
	}
}
\proof{From propositions \ref{nijol}, \ref{rtor} and \ref{dist-fol}.}
So we see, in the case where the endomorphism $I$ denotes a riemannian almost product structure, that the Nijenhuis tensor contains two parts $L,L'$, measuring the failure of integrability in the two complementary distributions $\esD,\esD'$ respectively. We also
see that an equivalent treatment is to look at the torsion of the Vidal connection which also measures the failure of integrability
of the two complementary distributions associated with $I$. Here we manifestly see the splitting of riemannian almost product structures into three
different types characterized by different conditions on the Nijenhuis tensor.
\prop{nij-fol}{Let the triplet $(\ts,\tg,I)$ denote a riemannian almost product structure, let $N_I$ denote the Nijenhuis tensor associated with $I$,
	then $N_I$ characterizes three different types by
	\eqal{2}{
		N_I(X,Y)&=0&			\qquad	&\text{\bf doubly foliated}\\
		IN_I(X,Y)&=\pm N_I(X,Y)&	\qquad	&\text{\bf singly foliated}\\
		IN_I(X,Y)&\neq\pm N_I(X,Y)&	\qquad	&\text{no foliation}.
	}
}
We will see some examples involving the two types of foliated almost product structures later, but we will first examine what extra structure these two types
give us. We will start by introducing two new differential operators associated with an almost product structure $I$.
\defn{nybound}{Let $I$ be an almost product structure on a manifold $\ts$ with exterior derivative $\td$. Let furthermore $\td_I$ denote
	the exterior derivative associated with $I$ and define two new differential operators by
	\eqsnono{
		\id&:=&{1\fr 2}(\td+\td_\I)\\
		\nd&:=&{1\fr 2}(\td-\td_\I)
	}
	An equivalent definition is by the two projection operators defined by the endomorphism $\I$, $\ico:={1\fr 2}(1+I)$ and
	$\nco:={1\fr 2}(1-I)$, then the operators are simply $\id\equiv\td_\ico$ and $\nd\equiv\td_{\nco}$.
}
These differential operators will be of utmost importance in the case where we have a vanishing Nijenhuis tensor.
\prop{bound}{Let $I$ be an almost product structure on a manifold $\ts$, and $N_I$ the Nijenhuis tensor associated with $I$, then if $N_I=0$ the
	new differential operators defined in \ref{nybound} will be nilpotent and thus coboundary operators. The exterior algebra will
	become doubly graded with respect to these new coboundary operators.}
\proof{We know from lemma \ref{nijenigen} that $N_I=0\Rightarrow N_{\ico}=N_{\nco}=0$ why both $\id$ and $\nd$ are nilpotent. 
	They are thus coboundary operators.
	Because of the doubly foliated structure we know that they can be expressed locally by $d=dx^m\pa_m$ and $\nd=dy^{m'}\pa_{m'}$.}
We see that if the Nijenhuis tensor vanishes the new differential operators are in fact coboundary operators and the exterior algebra becomes
doubly graded under these two coboundary operators.
\defn{doublegradation}{Let the triple $(\ts,\tg,I)$ define a riemannian almost product structure, let $N_I=0$ and denote the set of doubly graded
	forms on $\ts$ by $\O^{p,q}=\O^{p,q}(\ts)$ characterized by
	\eqnono{
		\o={1\fr (p+q)!}\o_{m_1\ldots m_p m'_1 \ldots m'_q}(x,y)dx^{m_1}\wedge \ldots dx^{m_p}\wedge dy^{m'_1}\wedge \ldots dy^{m'_q}
	}
	where $\o\in \O^{p,q}$. We thus see that the new coboundary operators defined in \ref{nybound} have the following characteristics:
	\eqal{2}{
		\id:&\O\longmapsto \O,&\qquad \O^{p,q}&\longmapsto \O^{p+1,q}\\
		\nd:&\O\longmapsto \O,&\qquad \O^{p,q}&\longmapsto \O^{p,q+1}
	}
	and that the graded algebra of differential forms now becomes doubly graded. The coboundary operators trivially satisfy the relations
	\eqsnono{
		d^2&=&0,\\
		\nd{}^2&=&0,\\
		 \id\nd+\nd\id&=&0.
	}
}
We see that it is in complete analogy to the case of an almost complex structure, where the vanishing of the Nijenhuis tensor tells us that we
have a complex manifold which gives us a doubly graded exterior algebra under holomorphic and anti-holomorphic coordinates. In this case though
we have a splitting which looks topologically like a product, taken into account that the almost product structure $I$ defines an Ehresmann foliation. 
This requirement is just that taking any curve in a leaf of the foliation $\esF$ and lifting it to another leaf by following only normal directions,
the quotient of their respective lengths shall exist. This is the same as saying that the curves do not shrink to zero or blow up to infinity
as we lift them by normal curves. There do exist foliations which have these types of singularities, also called Reeb components, see \cite{Rov98}.
In the case of a riemannian almost product structure defining a Ehresmann foliation this amounts to saying that we are assured that the induced metrics
on the two complementary distributions exist and are non-singular. So letting $I$ define an Ehresmann foliation with vanishing Nijenhuis tensor
it follows that the universal covering space splits to a topological product, 
$\ul{\tilde{\ws}}=\tilde{\ws} \times \tilde{\ws}'$, where the tilde denotes the universal covering space and the product is in the topological sense, 
see \cite{BlHe83,BlHe84}.
In an analogous way to the complex case, we also get  a splitting of the cohomology groups under these two new coboundary operators and the double gradation.
\thm{cohom}{Let $\ts$ be a manifold, let $I$ be an almost product structure on $\ts$ that defines an Ehresmann connection, then the vanishing of the Nijenhuis
	tensor implies that the de Rahm cohomology groups on $\ts$ splits like
	\eqnono{
		H^p(\dsR)=\underset{p=k+l}{\oplus}H^{k,l}(\dsR)
	}
	
}
\proof{See \cite{BlHe83}}
In the case of theorem \ref{cohom} we see that the basic cohomology groups map isomorphically into these doubly graded cohomology groups.
We have in this case $H^p_{B_{\esF}}=H^{0,p}$.
Let us finally list some local properties of the different tensors involved. We put them in a proposition
but the proof will be immediate. 
\prop{typesofI}{Let the triplet $(\ts,\tg,I)$ define a riemannian almost product structure, then we have three basic types defined by the Nijenhuis tensor.
	We will see how the local structure of the tensors involved look.
	\rnum{
		\item[(i)]	{\bf doubly foliated} $\Leftrightarrow$ $N_I=0$.\\
				In this case, where the Nijenhuis tensor of the almost product structure vanishes, 
				we have a doubly graded tensor algebra. We can therefore write the oriented vielbeins on the form
				\eqal{2}{
					E_a&=E_a{}^m\pa_m,& \qquad 	E_{a'}&=E_{a'}{}^{m'}\pa_{m'},\\
					E^a&=dx^mE_m{}^a,& \qquad	E^{a'}&=dy^{m'}E_{m'}{}^{a'},
				}
				where of course $E_a{}^m,E_{a'}{}^{m'}, E_m{}^a,E_{m'}{}^{a'}$ are functions on $\ts$ satisfying
			$E_a{}^mE_m{}^b=\d_a{}^b,E_m{}^aE_a{}^n=\d_m{}^n,E_{a'}{}^{m'}E_{m'}{}^{b'}=\d_{a'}{}^{b'},E_{m'}{}^{a'}E_{a'}{}^{n'}=\d_{m'}{}^{n'}$. 
				The metric takes the form
				\eqsnono{
					\tg&=&\eta_{ab}E^aE^b + \eta_{a'b'}E^{a'}E^{b'}=\\
					&=&\ig_{mn}(x,y)dx^mdx^n+\nlg_{m'n'}(x,y)dy^{m'}dy^{n'}.
				}
				where we have used $\eta$ to stress that we can have any signature of the metric.
				We also see that the almost product structure takes the simple form
				\eqsnono{
					I&=&E^aE_a-E^{a'}E_{a'}=\\
					&=&dx^m\pa_m-dy^{m'}\pa_{m'},
				}
				so the Nijenhuis tensor vanishes. We also find the two associated boundary operators
				to be
				\eqsnono{
					d&=&dx^m{\pa \fr \pa x^m},\\
					d'&=&dy^{m'}{\pa \fr \pa y^{m'}}.
				}
		\item[(ii)]	{\bf singly foliated} $\Leftrightarrow$ $I N_I = \pm N_I$.\\
				In this case only one set of vielbeins associated to $I$ defines a foliation which we will take to
				be the unprimed set, {\it i.e.}, $IN_I=N_I$. The vielbeins can now be expressed in the form
				\eqal{2}{
					E_a&=E_a{}^m\pa_m,& \qquad 	E_{a'}&=E_{a'}{}^{m'}(\pa_{m'}+A_{m'}{}^m\pa_m),\\
					E^a&=(dx^m - dy^{m'}A_{m'}{}^m)E_m{}^a,& \qquad	E^{a'}&=dy^{m'}E_{m'}{}^{a'},
				}
				now additionally $A_{m'}{}^m$ are functions on $\ts$. It is convenient to introduce objects
				$D_{m'}:=\pa_{m'}+A_{m'}{}^m\pa_m$ and $\Pi^m:=dx^m - dy^{m'}A_{m'}{}^m$ such that the vielbeins instead 
				can be written in the simpler form
				\eqal{2}{
					E_a&=E_a{}^m\pa_m,& \qquad 	E_{a'}&=E_{a'}{}^{m'}D_{m'}\\
					E^a&=\Pi^mE_m{}^a,& \qquad	E^{a'}&=dy^{m'}E_{m'}{}^{a'}.
				}
				Now there will be no surprise that $D_{m'}$ in fact will be the covariant derivative in the example of
				foliations in principle bundles that we will see later. The metric takes the form
				\eqsnono{
					\tg&=&\eta_{ab}E^aE^b + \eta_{a'b'}E^{a'}E^{b'}=\\
					&=&\ig_{mn}(x,y)\Pi^m\Pi^n+\nlg_{m'n'}(x,y)dy^{m'}dy^{n'}.
				}
				where the non-integrability of the prime distribution makes itself manifest through the differentials $\Pi^m$.
				We find the almost product structure to be of the form
				\eqsnono{
					I&=&E^aE_a-E^{a'}E_{a'}=\\
					&=&\Pi^m\pa_m-dy^{m'}D_{m'}
				}
				why the associated Nijenhuis tensor fails to vanish but instead reads
				\eqsnono{
					-{N_I}_{m'n'}&=&\ico [D_{m'},D_{n'}]=\\
						&=&\pa_{m'}A_{n'}-\pa_{n'}A_{m'}+[A_{m'},A_{n'}].
				}
				It thus measures at what extent the prime distribution fails to be integrable. The two associated differential operators
				become
				\eqsnono{
					d&=&\Pi^m\pa_m,\\
					d'&=&dy^{m'}D_{m'}.
				}
				Let $I$ define an Ehresmann connection, and thus a fibration. Denote it by 
				$0 \rightarrow \ws_{\esF} \rightarrow \ts \rightarrow \ws'_{\esD} \rightarrow 0$, where $\ws_{\esF}$ 
				is the leaf of the foliation and $\ws'_{\esD}=\ts/\ws_{\esF}$ is the leafspace. Let further 
				$\s$ be a section of the leafspace in $\ts$, then the covariant derivative on
				the leafspace is simply $D'=\s^*\td=\s^*d'=d'|_\s$. The curvature of this covariant derivative is nothing but the Nijenhuis
				tensor.
		\item[(iii)]	{\bf no} foliation $\Leftrightarrow$ no condition.\\
				In this case we have no foliation and thus the sets of vielbeins will none be of a simple form but both needs
				to be expressed in terms of both $\pa_m$ and $\pa_{m'}$. This case will be of no interest to us as we practically get
				no extra structure of importance.
	}
}
%%%%%%%%%%%%
%
% Subsubsection: Classes from Jordan tensor.
%
%%%%%%%%%%%%
\subsubsection{The classes defined by the Jordan tensor}

We will here proceed to get the extra structure to a riemannian almost product structure by looking at the Jordan tensor. If we put everything we
have regarding the Jordan tensor together we end up with the theorem.
\thm{GcoG}{Let the triplet $(\ts,\tg,I)$ define a riemannian almost product structure, let $\esD,\esD'$ be the associated distributions and let
	$L,L'$ be the skew tensors of the deformation, then the first type of almost product structure corresponding to 
	a doubly foliated manifold can be seen by the following equivalent statements,
	\eqal{2}{
		(i)&&\qquad	&M_I=0,\\
		(ii)&&\qquad	&K=0,\quad K'=0,\\
		(iii)&&\qquad	&\rdel\quad \text{is metric},\\
		(iv)&&\qquad	&\esD,\esD'\quad \text{are geodesic}.
	}
}
\proof{(i), (ii) and (iv) is clear from proposition \ref{almostjordan2} and definition \ref{umg}, now we need to prove (iii), that is we need to prove that
	\eqal{1}{
		\tg(B(X,Y),Z)+&\tg(Y,B(X,Z))=\\
		=&{1\fr 4}(\tg((I \tdel_Y I + \tdel_{IY})(X),Z)+\tg(Y,(I \tdel_Z I + \tdel_{IZ})(X)))=\\
		=&-{1\fr 4}(\tg(X,(I \tdel_Y I - \tdel_{IY})(Z))+\tg((I \tdel_Z I - \tdel_{IZ})(Y),X))=\\
		=&-{1\fr 4}\tg(X,H_I(Y,Z)+H_I(Z,Y))=\\
		=&-{1\fr 8}\tg(X,M_I(Y,Z)),
	}
	and the equivalence is clear.
}
We see that we have a similar structure as in the case for the Nijenhuis tensor. Now the Jordan tensor measures whether the two complementary
distributions are geodesic or not while the Nijenhuis tensor measured whether they were integrable. We will soon proceed to split the Jordan tensor
further and look at the traceless and trace parts of it to divide up the classes further, but first we will look at the special case when the
almost product structure $I$ is covariantly constant, as we will see a typical analogue to the complex case.
\thm{equiv}{Let the 3-tuple ($\ts$, $\tg$, $\I$) define a riemannian almost product structure with Levi--Civita connection $\tdel$. Let
	$\adel$ denote the adapted connection and $\rdel$ the Vidal connection
	then the following equivalence holds
	\eqnono{
		\tdel I=0\quad\Leftrightarrow\quad\rdel=\adel=\tdel
	}
}
\proof{Immediate from definitions \ref{adap} and \ref{rdel}.}
The first obvious consequence of this is that the Nijenhuis tensor vanishes. Note that in the case of a K\"ahler manifold we know that
as the almost complex structure, $J$, is covariantly constant, {\it i.e.}, $\tdel J=0$, $J$ is also integrable.
\cor{ni}{Let the 3-tuple ($\ts$, $\tg$, $\I$) define a riemannian almost product structure and let $\tdel$ be the Levi--Civita connection then
	\eqnono{
		\tdel I=0\quad\Rightarrow\quad N_I=0 
	}
}
\proof{From \ref{equiv} and proposition \ref{del}.}
We also know from the complex case that K\"ahler implies reduction of the holonomy groups so it is no surprise that we find it in the case
of a covariantly constant almost product structure to.
\cor{equivagain}{Let the 3-tuple ($\ts$, $\tg$, $\I$) define a riemannian almost product structure. If the adapted connection and the Vidal connection
	are Levi--Civita then the holonomy group splits as
	\eqnono{
		O(m)=O(k)\times O(k')
	}
	which follows from the splitting of the Lie algebra of the connection
	\eqnono{
		\goo(m)=\goo(k)\oplus\goo(k')
	}
}
\proof{From proposition \ref{del}.}
In the case of a K\"ahler manifold we know that the holonomy group reduces to $U(m)\subset O(2m)$ where $U(m)$ is a subgroup of the generic holonomy group
$O(2m)$ of a $2m-$dimensional manifold, while in the case of the covariantly constant almost product structure we get the subgroup 
$O(k) \times O(m-k)$ instead of the generic holonomy group $O(m)$. From what we have seen the case of a covariantly constant almost product
structure tells us that the universal covering space in fact is a product manifold.
\thm{product}{Let the 3-tuple ($\ts$, $\tg$, $\I$) define a riemannian almost product structure. If now the Vidal connection $\rdel$ is
	Levi--Civita ({\it i.e.}, metric and torsionless), then $\tilde{\ts}$, the universal covering space of $\ts$, is a product manifold.}
\proof{From proposition \ref{GcoG} plus the fact that it is a topological product from the vanishing of the Nijenhuis tensor.}
Now we will continue to split the Jordan tensor into its traceless and trace parts to get the four classes of distributions in the geometric sense, namely
geodesic, umbilic, minimal, and the last with no condition. So as we saw in definition \ref{umg} we have eight different classes of a distribution and
now in the case of a almost product structure which leaves us with two complementary distributions we thus get 64 different classes. Now
it is immediate that it does not matter which we call complementary of the two distributions so we have in fact only 36 different classes, see \cite{Nav84}. 
\prop{apclasses}{Let the triplet $(\ts,\tg,I)$ be an riemannian almost product structure. We then
	have the following 36 different classes
	\eqnono{\begin{tabular}{c|c|c|c|c|c|c|l}
		Classes&$L$&$W$&$\k$&$L'$&$W'$&$\k'$&Name\\\hline\hline
		(GF,GF)&x&x&x&x&x&x&Local product\\\hline
		(GF,UF)&x&x&x&x&x& &Twisted product\\
		(GF,MF)&x&x&x&x& &x&\\
		(GF,F) &x&x&x&x& & &\\\hline
		(UF,UF)&x&x& &x&x& &Double twisted product\\
		(UF,MF)&x&x& &x& &x&\\
		(UF,F) &x&x& &x& & &\\\hline
		(MF,MF)&x& &x&x& &x&\\
		(MF,F) &x& &x&x& & &\\\hline
		(F,F)  &x& & &x& & &\\\hline\hline
		(GF,GD)&x&x&x& &x&x&Riemannian foliation\\
		(UF,GD)&x&x& & &x&x&Riemannian foliation\\
		(MF,GD)&x& &x& &x&x&Riemannian foliation\\
		(F,GD) &x& & & &x&x&Riemannian foliation\\\hline
		(GF,UD)&x&x&x& &x& &\\
		(UF,UD)&x&x& & &x& &\\
		(MF,UD)&x& &x& &x& &\\
		(F,UD) &x& & & &x& &\\\hline
		(GF,MD)&x&x&x& & &x&\\
		(UF,MD)&x&x& & & &x&\\
		(MF,MD)&x& &x& & &x&\\
		(F,MD) &x& & & & &x&\\\hline
		(GF,D) &x&x&x& & & &\\
		(UF,D) &x&x& & & & &\\
		(MF,D) &x& &x& & & &\\
		(F,D)  &x& & & & & &\\\hline\hline
		(GD,GD)& &x&x& &x&x&\\
		(GD,UD)& &x&x& &x& &\\
		(GD,MD)& &x&x& & &x&\\
		(GD,D) & &x&x& & & &\\\hline
		(UD,UD)& &x& & &x& &\\
		(UD,MD)& &x& & & &x&\\
		(UD,D) & &x& & & & &\\\hline
		(MD,MD)& & &x& & &x&\\
		(MD,D) & & &x& & & &\\\hline
		(D,D)  & & & & & & &\\\hline
		\end{tabular}
	}
}
The structure added to the various tensors in some of the different classes will be put as a proposition. Again the proof is immediate.
\prop{metricstructure}{Let the triplet $(\ts,\tg,I)$ define a riemannian almost product structure, let the Nijenhuis tensor
	define the three different types of almost product structures as in \ref{typesofI}, then we have additionally the following
	examples of classes in various types
	\rnum{
		\item[(i)]	{\bf doubly foliated} $\Leftrightarrow$ $N_I=0$.\\
				In this type we have 10 different classes as seen in proposition \ref{apclasses}. We will take a closer
				look at the local structure of the metric in some of these classes.
				\eqal{2}{
					(GF,GF):&&\qquad	\tg=&\ig_{mn}(x)dx^mdx^n+\nlg_{m'n'}(y)dy^{m'}dy^{n'}\\
					(GF,UF):&&\qquad	\tg=&\ig_{mn}(x)dx^mdx^n+\l'(x,y)\nlg_{m'n'}(y)dy^{m'}dy^{n'}\\
					(UF,UF):&&\qquad	\tg=&\l(x,y)\ig_{mn}(x)dx^mdx^n+\l'(x,y)\nlg_{m'n'}(y)dy^{m'}dy^{n'}\\
				}
				These are the classes referred to as {\bf local product}, {\bf twisted product} and {\bf doubly twisted product} respectively.
				In all these cases we know that $W=0,W'=0$, but in the cases where we have twisted products the mean curvature
				does not vanish. In fact we have
				\eqal{3}{
					(GF,UF):&&\qquad	\k=&0&\quad,		\k'=&-{k'\fr 2}{\l'}^{\minus 1}d\l'(x,y),\\
					(UF,UF):&&\qquad	\k=&-{k\fr 2}{\l}^{\minus 1}d'\l(x,y),&\quad 	\k'=&-{k'\fr 2}{\l'}^{\minus 1}d\l'(x,y).
				}
				These two classes are called {\bf warped product} and {\bf doubly warped product} respectively when the conformal factors
				$\l,\l'$ only depends on the coordinates $y^{m'},x^m$ respectively. This gives us
				\eqal{2}{
					(GF,UF):&&\qquad	\tg=&\ig_{mn}(x)dx^mdx^n+\l'(x)\nlg_{m'n'}(y)dy^{m'}dy^{n'},\\
					(UF,UF):&&\qquad	\tg=&\l(y)\ig_{mn}(x)dx^mdx^n+\l'(x)\nlg_{m'n'}(y)dy^{m'}dy^{n'},\\
				}
				and the mean curvatures now become basic 1-forms and take the form
				\eqal{3}{
					(GF,UF):&&\qquad	\k=&0,&\quad		\k'=&-{k'\fr 2}{\l'}^{\minus 1}d\l'(x),\\
					(UF,UF):&&\qquad	\k=&-{k\fr 2}{\l}^{\minus 1}d'\l(y),&\quad 	\k'=&-{k'\fr 2}{\l'}^{\minus 1}d\l'(x).
				}
				We will later see that this case is the interesting case of the present brane solutions in M-theory.
		\item[(ii)]	{\bf singly foliated} $\Leftrightarrow$ $I N_I = \pm N_I$.\\
				As in \ref{typesofI} we will look at the case where $IN_I=N_I$ and find that some structure is inherited
				from the doubly foliated case. Of this type we have 16 classes of which we will list some examples.
				\eqal{2}{
					(F,GD):&&\qquad	\tg=&\ig_{mn}(x,y)\Pi^m\Pi^n+\nlg_{m'n'}(y)dy^{m'}dy^{n'},\\
					(F,UD):&&\qquad	\tg=&\ig_{mn}(x,y)\Pi^m\Pi^n+\l(x,y)\nlg_{m'n'}(y)dy^{m'}dy^{n'}.\\
				}
				The first of these are called {\bf riemannian foliations} which are characterized by the complementary
				distribution being geodesic. Or to put it in the classification scheme, $(*F,GD)$. In this case
				the metric $\tg$ is said to be bundle-like and we have $\Lie_X\nlg=0$. So the vectors of the foliation
				are isometries of the complementary metric, $\nlg$. In the second case we see that we get a non-vanishing
				mean curvature form for the complementary distribution
				\eqnono{
					\k'_m=-{k'\fr 2}{\l'}^{\minus 1}\pa_m\l'.
				}
				If we now let the foliation be geodesic we get
				\eqal{2}{
					(GF,GD):&&\qquad	\tg=&\ig_{mn}(x)\Pi^m\Pi^n+\nlg_{m'n'}(y)dy^{m'}dy^{n'},\\
					(GF,UD):&&\qquad	\tg=&\ig_{mn}(x)\Pi^m\Pi^n+\l(x,y)\nlg_{m'n'}(y)dy^{m'}dy^{n'},\\
				}
				where additionally $A_{m'}{}^a=A_{m'}{}^a(y),C_{a(bc)}=0$. This is the case where the leaves of the foliation are a
				Lie group for instance. We will later see that a principal bundle lies in the first of these classes.
				Furthermore, we could let the foliation become umbilic and get an analogue of the type one case, but we will restrict to the case
				where the metric takes the form
				\eqal{2}{
					(UF,GD):&&\qquad	\tg=&\l(y)\ig_{mn}(x)\Pi^m\Pi^n+\nlg_{m'n'}(y)dy^{m'}dy^{n'}.\\
				}
				We will see that in Kaluza--Klein theory this is the case when introducing the scalar field $\l=e^{-2\phi}$ 
				which measures the radius of the gauge group. Here we get
				\eqnono{
					\k_{m'}=-{k\fr 2}\l^{\minus 1}\pa_{m'}\l=k\pa_{m'}\phi.
				}
		\item[(iii)]	{\bf no} foliation $\Leftrightarrow$ no condition.
				As this case is rather uninteresting we will only say that in the case of a geodesic distribution
				the extrinsic curvature vanishes which can be viewed in the form
				\eqal{2}{
					(GD):&&\qquad	K_{abc'}=C_{c'(ab)}=0.
				}
	}
}
We can now use this formalism to study the structure of for instance the brane solutions of M-theory. The following example
tells how the different tensors look and what they say.
\exam{m2brane}{{\bf M2-brane}\\
	In the M2-brane case we have the solution to the equations of motion for the metric in the form \cite{BeSeTo87,Du96}
	\eqnono{
		\tg=\Delta^{\minus {2\fr 3}}(y)\eta_{mn}dx^mdx^n+\Delta^{{1\fr 3}}(y)\d_{m'n'}dy^{m'}dy^{n'}
	}
	where
	\eqnono{
			\Delta(y)=1+({a\fr \r(y)})^6,\qquad \r(y)=\sqrt{\d_{m'n'}y^{m'}y^{n'}}.
	}
	and $\r=0$ is the horizon of the brane not the core. We find the corresponding vielbeins
	\eqal{2}{
		E_a=&\Delta^{\minus {1\fr 3}}\d_a{}^m\pa_m,&\qquad	E_{a'}=&\Delta^{{1\fr 6}}\d_{a'}{}^{m'}\pa_{m'},\\
		E^a=&\Delta^{1\fr 3}dx^m\d_m{}^a,&\qquad		E^{a'}=&\Delta^{\minus {1\fr 6}}dy^{m'}\d_{m'}{}^{a'},
	}
	which we use to derive the almost product structure which splits the tangent bundle accordingly
	\eqsnono{
		I&=&E^aE_a-E^{a'}E_{a'}=\\
		&=&dx^m\pa_m-dy^{m'}\pa_{m'}.
	}
	By definition, $I^2=1$, and we see that not only the brane is integrable but also the complementary distribution
	associated with $I$. Accordingly the Nijenhuis tensor vanishes,
	\eqnono{
		N_I=0.
	}
	So we see that this typical solution is a doubly foliated manifold in the class (UF,GF) and additionally it is spherical
	why the metric is nothing but a warped-product. We get the mean curvature
	\eqnono{
		\k=\Delta^{\minus 1}d'\Delta,
	}
	and as $W$ vanishes we see that generating translations radially from the brane by the vector $\pa/\pa\r$ we generate conformal
	transformations on the brane.\\

	{\bf M5-brane}\\
	In the M5-brane case we will look at two types of solutions, the first one of which is the ordinary with no field excitations
	on the brane \cite{Gu92,Du96}, the second where we have excited the anti self-dual tensor field on the brane found recently \cite{Oss}. The
	first solution of the metric looks like
	\eqnono{
		\tg=\Delta^{\minus {1\fr 3}}(y)\eta_{mn}dx^mdx^n+\Delta^{{2\fr 3}}(y)\d_{m'n'}dy^{m'}dy^{n'}
	}
	where
	\eqnono{
			\Delta(y)=1+({a\fr \r(y)})^3,\qquad \r(y)=\sqrt{\d_{m'n'}y^{m'}y^{n'}}.
	}
	The metric and thus the vielbeins look very similar to the M2-brane case
	\eqal{2}{
		E_a=&\Delta^{\minus {1\fr 6}}\d_a{}^m\pa_m,&\qquad	E_{a'}=&\Delta^{{1\fr 3}}\d_{a'}{}^{m'}\pa_{m'},\\
		E^a=&\Delta^{1\fr 6}dx^m\d_m{}^a,&\qquad		E^{a'}=&\Delta^{\minus {1\fr 3}}dy^{m'}\d_{m'}{}^{a'},
	}
	Again we get $I=dx^m\pa_m-dy^{m'}\pa_{m'}$ and $N_I=0$ but now of course with a different number of $x$ and $y$ directions.
	So we see also in this case that we have a doubly foliated solution of the type $(UF,GF)$. Notable is that this implies that
	the antisymmetric tensor fields which where a basic form now lies in the graded cohomology group
	\eqnono{
		H\in H^{0,4}.
	}
	The mean curvature is the same as in the M2-brane case
	\eqnono{
		\k=\Delta^{\minus 1}d'\Delta,
	}
	but now with a different function $\Delta$ of course.
	In the other solution with the tensor field excited the metric looks like
	\eqnono{
		\tg=(\Delta_+\Delta_-)^{\minus 1/6}\left[\left({\Delta_+\fr \Delta_-}\right)^{1/2}dx_-^2+\left({\Delta_-\fr \Delta_+}\right)^{1/2}dx_+^2\right]
			+(\Delta_+\Delta_-)^{1/3}dy^2
	}
	where $\Delta_+=\Delta+\nu, \Delta_-=\Delta-\nu$ and $\Delta$ is as before. Here we have yet another almost product structure lying in the brane denoted
	$q$ which squares to one. In this case we find that $W$ does not vanish anymore but will in fact be
	\eqnono{
		\li W={1\fr 4}({1\fr \Delta_+}- {1\fr \Delta_-})\li qd'\Delta
	}
	and the mean curvature will read
	\eqnono{
		\k={1\fr 2}({1\fr \Delta_+}+ {1\fr \Delta_-})d'\Delta
	}
	so we see that we have a solution in the class $(F,GF)$. Interesting to see is that the new almost product structure in the brane, $q$,
	defines three new preferred directions which in some sense can be seen as a membrane, see \cite{Oss} for further information.
	Here we just state the utmost importance in studying several almost product structures on a manifold as these would result in multi brane
	configurations. Interesting would be to see what conditions would be implied on these almost product structure if we furthermore require
	that these associated brane configurations would solve the equations of motions.
}
Now as we said earlier the structure of the Nijenhuis tensor took such a form that we suspected that it measured the curvature
of fibrations. It will be clear from the next example, where we look at principal bundles, that this is truly a fact. We will
also see that Nijenhuis tensor measures the field strength in Kaluza--Klein theories.
\exam{liefoliation}{Let $P(\ws,G)$ be a principal bundle with base space $\ws$ and fiber $G$. Let furthermore $\td$ denote the
	exterior derivative in the total space, $T_a$ denote the generators of the Lie algebra $\gog$ associated to the Lie group $G$, fulfilling
	the algebra $[T_a,T_b]=f_{ab}{}^cT_c$ and normalized like $\tr (T_a T_b)=\d_{ab}$. The vielbeins can be expressed as
	\eqal{2}{
		E^aT_a=&g^{\minus 1}\td g+g^{\minus 1} A g,& \qquad	E^{a'}=&dy^{m'}E_{m'}{}^{a'},\\
		E_a=&R_a,& \qquad					E_{a'}=&E_{a'}{}^{m'}(\pa_{m'}-A_{m'}{}^aR_a)
	}
	where $g^{\minus 1}\td g(L_a)=T_a, g^{\minus 1}\td g(R_a)=Ad_{g^{\minus 1}}T_a$ and $R_a$ and $L_a$ are the right and left invariant
	vector fields on $G$ respectively. 
	Here we have done the split $T_u P=V_u P \oplus H_uP$ where the vertical subbundle is spanned by $E_a$ and the horizontal by
	$E_{a'}$. 
	We define the connection $\o:=E^aT_a$ \cite{EgGiHa80,Is89} and write the Lie algebra valued curvature form as
	\eqnono{
		\O:=\td \o + \o \wedge \o
	}
	We now know \cite{Nak90} that taken two vectors $X,Y\in TP$ we get
	\eqnono{
		\O(X,Y)=-\o([X_H,Y_H]).
	}
	This tells us that if we expand the connection to $\tilde{\o}(g,y):=l_{g*}\o=E^aL_a$ now giving $\tilde{\o}(X)=X_V$ instead
	of pushing the vector back to the Lie algebra we see that, defining $\tilde{\o}':=1_P-\tilde{w}$ and $I=\tilde{\o}-\tilde{\o}'=2\tilde{\o}-1_P$,
	we get from lemma \ref{nijenigen}
	\eqnono{
		\tilde{\O}(X,Y)={1\fr 4}N_I(X,Y)
	}
	or ${1\fr 4}\o(N_I(X,Y))=\O(X,Y)$. So we see that indeed the field strength of gauge theory is a special case of the Nijenhuis tensor
	only valid when the fiber is a gauge group. We can also easily see that this is nothing but a foliation of class (GF,GD). This due to the fact
	that by definition we have $H_{ug}P=r_{g*}H_uP$ which implies that $A=A(y)$ implying $GF$ and $E_{a'}{}^{m'}=E_{a'}{}^{m'}(y)$ implying $GD$.
	
	In the case of algebraic gauge \cite{LaMa90} we have the short exact sequence $0 \rightarrow A \rightarrow^i E \rightarrow^\pi B \rightarrow 0$
	where $A$ is the fiber $E$ the total space and $B$ the base manifold all being algebras. We have a connection on $B$ denoted $\r$ such that
	\eqnono{
		\r:B \longmapsto E,\qquad \pi \circ \r=1_B.
	}
	Equivalently we can look at a connection in $E$ instead and denote it by $\o$ now satisfying
	\eqnono{
		\o:E\longmapsto A,\qquad \o\circ i=1_A.
	}
	We have the immediate relation between the two connections
	\eqnono{
		\o=1_E-\r\circ \pi
	}
	satisfying $\o^2=\o$. The curvature of these two connections are defined for $X,Y\in \L^1_B$ and $\ul{X},\ul{Y}\in\L^1_E$ by
	\eqsnono{
		F(X,Y)&:=&\r([X,Y])-[\r(X),\r(Y)],\\
		\O(\ul{X},\ul{Y})&:=&F(\pi\ul{X},\pi\ul{Y}).
	}
	Now let $\o'=1_E-\o=\r\circ \pi$ and $I=\o-\o'=1_E-2\r\circ \pi$ then the curvature $\O$ is nothing but the Nijenhuis tensor or
	\eqnono{
		\O(\ul{X},\ul{Y})={1\fr 4}N_I(\ul{X},\ul{Y})
	}
	which follows directly when expressing $\O$ in terms of $\o$, see \cite{LaMa90}.
}
Next example will be that of Kaluza-Klein theory.
\exam{Kaluza--Klein}{We can extend the above example to the case of Kaluza--Klein theory where the fiber needs not be the gauge group itself
	but rather having the gauge group as isometry group \cite{DuNiPo85}. Again we split the space as a fibration looking first at the $(GF,GD)$ case. 
	Let $\ts$ denote the total space and $\ws$ the base space. Let furthermore $H$ denote the fiber which and let $(x^m,y^{m'})$ be local coordinates
	such that $\{\pa_m\}$ spans the foliation, $H$, locally and thus write our adapted frames
	\eqnono{
		\hat{E}_a=E_a{}^m\pa_m,\qquad \hat{E}_{a'}=E_{a'}{}^{m'}\pa_{m'}+A_{a'}{}^{m}\pa_m
	}
	Let $K_i$ be the set of Killing vectors of the fiber fulfilling the algebra
	\eqnono{
		[K_i,K_j]=f_{ij}{}^kK_k
	}
	where of course $K_i=K_i{}^a(x)E_a$, so we can express $A=E^{a'}A_{a'}^iK_i$. We also require that $A^i=A^i(y)$ and that $f_{i(jk)}$=0 (see 
	proposition \ref{metricstructure}). The inverse vielbeins now read
	\eqnono{
		\hat{E}^a=E^a-A^iK_i^a,\qquad \hat{E}^{a'}=E^{a'},
	}
	from which we derive the metric of the total space
	\eqal{1}{
		\tg=&\ig+\nlg=\eta_{ab}\hat{E}^a\hat{E}^b+\eta_{a'b'}\hat{E}^{a'}\hat{E}^{b'}=\\
			=&\eta_{ab}(E^a-A^iK_i^a)(E^b-A^jK_j^b)+\eta_{a'b'}E^{a'}E^{b'}.
	}
	We also use the set of vielbeins to form the almost product structure which splits the space according to the fibration. As it is a fibration
	this almost product structure can be seen as an Ehresmann connection on $\ts$.
	\eqnono{
		I=\hat{E}^a\hat{E}_a-\hat{E}^{a'}\hat{E}_{a'}=E^aE_a-E^{a'}E_{a'}-2A^iK_i
	} 
	Of course $I^2=1$, and if $X,Y\in\L^1$ we have from lemma \ref{nijenigen}
	\eqnono{
		-N_I(X,Y)=\ico [ \nco X, \nco Y].
	}
	Let $\hat{E}_{m'}=\pa_{m'}+A_{m'}^iK_i=:D_{m'}$ and note that $\ico K_i=K_i$ then
	we again see that the Nijenhuis tensor measures the curvature
	\eqal{1}{
		-{N_I}_{m'n'}=&\ico [ \hat{E}_{m'},\hat{E}_{n'} ]=\\
			=&\ico [ \pa_{m'}+A_{m'}^iK_i,\pa_{n'}+A_{n'}^jK_j]=\\
			=&(\pa_{m'}A_{n'}^i-\pa_{n'}A_{m'}^i+f_{jk}{}^iA_{m'}^jA_{n'}^k)K_i=\\
			=&F^i_{m'n'}K_i.
	}
	From this analysis it is clear that we can extend this fibration to any case in the classification scheme, {\it i.e.}, $(*F,*D)$, where 
	$*$ means $G,U,M$ or nothing. So for example we can extend the theory to the $(UF,GD)$ case where we have added one additional
	factor which can conformally transform the fiber as we move along the base space. So by letting $\phi=\phi(y)$ be a scalar field, 
	often referred to as the dilaton, and rescale the vielbein as
	\eqnono{
		\tilde{E}^a:=e^{-\phi}\hat{E}^a
	}
	then the mean curvature becomes
	\eqnono{
		\k_{m'}=-{1\fr 2}k\l^{\minus 1}\pa_{m'}\l=k\pa_{m'}\phi,
	}
	telling us that the fiber now is umbilic instead or that movement on the base space generates conformal transformations on the fiber.
	The dilaton can now be seen as measuring the radius of the fiber.
	We could of course go further by relaxing the conditions on the foliation and the complementary distribution further. Notable is
	that if we want to relax the condition on the fiber further we will break some of the isometries so the Killing vector algebra will reduce.
	We also note that any further relaxation of these kinds will not alter the Nijenhuis tensor and thus not the gauge field strength either.
}
%%%%%%%%%%%%%%%%%%%%%%%%%%%%%%%%%%%%
%
% Section: Conclusions and outlook
%
%%%%%%%%%%%%%%%%%%%%%%%%%%%%%%%%%%%%
\section{Conclusions and outlook}

We have demonstrated that we can express gauge theory, Kaluza--Klein theory and brane theory as special cases of almost product manifolds.
We have also seen that the Nijenhuis tensor of certain almost product structures measures integrability which in gauge theories and
Kaluza--Klein theories implies that the field strength measures the non-integrability of the complementary distribution to the foliation
associated with the fiber (as we found them to be equal). Now there are lots of things that could be investigated further, one
is how this almost product structure appears in the Clifford algebra. Another thing is to generalize all this to superspace, as
we know from the embedding formalism \cite{HoSeWe97,BaPaSoToVo95,Oss2} that a simple constraint gives us the right multiplets and brings
us on shell. In a forthcoming paper \cite{Mig}, we will show how flat superspace can be seen in this formalism. But the most interesting
continuation  would of course be to find new solutions that are non trivial and maybe even show the existence of solutions
in which the leafspace is non-commutative.
\vskip1cm
\ul{Acknowledgments}:
The author would like to thank Martin Cederwall for all help he has given with this paper and also Niclas Sandstr\"om for discussions.
%$${\scr{ABCDEFGHIJKLMNOPQRSTUVWXYZ}}$$
%$${\cal{ABCDEFGHIJKLMNOPQRSTUVWXYZ}}$$
%$${\EuScript{ABCDEFGHIJKLMNOPQRSTUVWXYZ}}$$
%\bibliographystyle{plain}
%\bibliography{paper3}

\end{document}